\documentclass[twocolumn]{article}
\usepackage{breqn}

\newcommand{\text}[1]{#1}
\newcommand{\sqrtpw}[1]{\left(#1\right)^{1/2}}
\newcommand{\zfrac}[2]{\Big(#1\Big)/\Big(#2\Big)}
\newcommand{\bed}{\begin{dmath}}
\newcommand{\eed}{\end{dmath}}
\newcommand{\be}{\begin{equation}}
\newcommand{\ee}{\end{equation}}
\newcommand{\tmp}{aux}
\newcommand{\Nac}{N_a\!}
\newcommand{\Nbc}{N_b\!}
\newcommand{\Ncc}{N_c\!}
\newcommand{\Ndc}{N_d\!}
\newcommand{\Nec}{N_e\!}
\newcommand{\Nfc}{N_f\!}
\newcommand{\Ngc}{N_g\!}

\begin{document}
\onecolumn

\title{$\overline{SO}(5)$ Clebsch-Gordan coefficients involving the
$14$-dimensional representation}
\author{Igor Salom and Djordje \v Sija\v cki \\
Institute of Physics, University of Belgrade, PO Box 57, 11001
Belgrade, Serbia}

\maketitle

\begin{abstract}
Analytic expressions for the Clebsch-Gordan (CG) coefficients of
the $\overline{SO}(5)$ group that involve the $14$-dimensional representation
can be found in an old paper of M. K. F. Wong. A careful analysis
yields that roughly 30\% of the coefficients given in that paper
are wrong. The correct analytic expressions for all $\overline{SO}(5)$ group
CG coefficients containing the $14$-dimensional representation are
obtained.
\end{abstract}

\section{Introduction}

The $SO(5)$ group, i.e. its covering group $\overline{SO}(5) \simeq Spin(5)$,
and its Lie algebra $so(5)$ have significant applications in
various fields of physics, for example, in atomic, molecular and solid state
physics, nuclear physics, grand unified theories of elementary particles,
higher dimensional gravity theories, and the physics of $pD$-branes. The
analytic expressions of the $\overline{SO}(5)$ group Clebsch-Gordan (CG)
coefficients,
containing the $14$-dimensional irreducible representation, apply mutatis
mutandis to various relevant cases in all of these fields of physics. The CG
coefficients are required whenever the symmetries of the physical system are
used for the evaluation of matrix elements of tensor operators. These tensor
operators can vary from diverse Hamiltonian pieces in molecular and nuclear
physics to gauge potentials, field strengths, connection coefficients,
curvatures and torsions in the physics of particles, fields and gravity. For
the sake of brevity, we mention below only a few examples.

The $\overline{SO}(5)$ symmetry appears as a hidden symmetry of the spin
$\frac{3}{2}$ atomic systems regardless of the dimensionality,
lattice geometry and impurity potentials. It plays the role of the
$SU(2)$ symmetry in spin $\frac{1}{2}$ systems. \cite{R1, R2}. The
$SO(5)$ group is utilized in nuclear physics as a dynamical and/or
spectrum generating group. For instance, in an $SO(3)$ basis
\cite{R3}, it plays a prominent role in the classification of
states of the nuclear collective model and the Interacting Boson
Model \cite{R4, R5}. Moreover, the $SO(5)$ dynamical symmetry
group plays a prominent role in the subject of exactly solvable
nuclear models with non-degenerate single-particle energies
\cite{R6}. Superconductivity in the cuprates is characterized by
its proximity to the antiferomagnetic phase with a tendency
towards superconductivity (SC) in competition with that towards
antiferromagnetism (AF). The high temperature superconductivity
(HTSC) theory, which emphasizes this aspect, is a unified SC-AF
theory based on the $SO(5)$ symmetry of a $5$-component
``superspin'' (with 3 components of the antiferromagnetic, and 2
for the superconducting order parameter) \cite{R7}. Due to a close
relation of the $so(5)$ algebra to the algebras of the (anti)De
Sitter and Lorentz-like groups in $5$ dimensions, many results
obtained for the $SO(5)$ group can be
straightforwardly conveyed and applied to the cases of these
groups. The $SO(5)$ and/or $\overline{SO}(5)$ groups CG coefficients, in
particular those coupling coefficients for combining the
$14$-dimensional and an arbitrary irreducible representation to
other irreducible representations, are of essential importance for
an explicit construction of the unitary irreducible representations of
the $SL(5,R)$ and/or $\overline{SL}(5,R)$ groups \cite{R8}. It turns out that
the noncompact $SL(5,R)$ generators can be constructed by generalizing
\cite{R9} the so called Gell-Mann formula \cite{R10} that expresses these
generators in terms of the compact and the corresponding Wigner-In\"on\"u
contracted noncompact generators. This construction utilizes
certain CG coefficients given below.
The $SL(5,R)$ and/or $\overline{SL}(5,R)$ groups are the homogeneous parts of
the gauge symmetries of the affine generalizations \cite{R11, R12} of the
(Kaluza-Klein-like) theories of gravity in $5D$ \cite{R13}. The symmetry
breaking to the Poincar\'e gauge gravity and evaluation of the thus generated
shear-connection mass requires knowledge of certain $SO(5)$ CG coefficients
as determined by the Higgs field choice. Moreover, in the quantum
field theory case, the interaction vertices of the infinite component
spinorial matter and the shear gauge potentials, $\bar\Psi \Gamma^{(ab)}_{\mu}
\Psi$ are given in terms of all
possible (infinitely many) spinorial CG coefficients of this work.
The spinorial irreducible representations of the $\overline
SL(5,R)$ in the $Spin(5)$ basis are vital for construction of
infinite component curved-spacetime Dirac-like equations and the
spinning $5D$-branes \cite{R14, R15}.

Numerical evaluation of the CG coefficients can be rather
straightforwardly performed using contemporary computers. However,
quite often it is very useful, even demanding, to have analytic
expressions of the CG coefficients, e.g when studying representations of
groups in the $SO(5)$ subgroup basis. These formulas allow one to
study certain expressions and their asymptotic behavior. Analytic
expressions for the $SO(5)$ ($Spin(5)$) CG coefficients (for
relations to the $SO(4) \simeq SU(2)\times SU(2)$ cf. \cite{R16})
involving a direct product of the $14$-dimensional, with an
arbitrary irreducible representation of the $SO(5)$ group were
published some time ago by Wong \cite{R17}. An attempt to make use
of the results of that paper  resulted in difficulties that turned
out to be due to numerous coefficients being erroneous (note that
computer capabilities were rather slim at the time the paper was
written). It turned out to be quite tedious and time consuming to
correct all explicitly presented expressions, as roughly 30\% of
the reduced CG coefficients, more precisely 34 out of 112, were
erroneous. Moreover, all other CG coefficients obtained from these
by making use of the symmetry properties bear the same destiny.

A partial list of various types of errors in Wong's paper
\cite{R17} is as follows: Eq. (7a) -- the factor $(a-b+c+d+2)$ is
to be replaced by $(a-b+c+d+1)$; Eq. (9a) -- the factor
$(a+a^2-b-2ab+b^2-c-c^2-d-d^2)$ is missing; Eqs. (6c), (8c) --
there should be $4b$ instead of $3b$ in the last bracket; Eq. (9d)
-- there should be $+2cd$ instead of $-2cd$ in the last bracket;
Eq. (12d) -- an overall sign of the coefficient should be +; Eq.
(13d) -- an overall sign should be + and the factor $(b - a + c +
d + 1)$ is to be replaced by $(b - a + c + d - 1)$; Eq. (2e) --
the factor $(a - b + c - d + 2)$ is to be replaced by $(a + b + c
- d + 2)$; Eq. (3e) -- the factors $(a - b + c - d + 3) (a - b + c
- d + 2)$ is to be replaced by $(a - b + c - d + 2) (a + b + c - d
+ 3)$; Eq. (5e) -- the factor $(a + b + c - d + 2)$ is to be
replaced by $(a + b + c - d + 1)$; Eq. (6e) -- the factor $((c +
1) (a + b + c + 3) +  d (d + 1))$ is to be replaced by $((c + 1)
(a + b + c + 3) - d (d + 1))$; Eq. (6f) -- the factor $(a + b + c
- d + 1)$ is to be replaced y $(a + b + c - d + 2)$; Eq. (8f) --
the factor $((d + 1) (a - b + c + 2) - c (c + 1))$ is to be
replaced by $((d + 1) (a - b + d + 2) - c (c + 1))$; Eq. (9f) --
the factor $((a + b)^2 - c^2 - d^2 - c - d + 2)$ is to be replaced
by $((a + b)^2 + 3 (a + b) - c^2 - d^2 - c - d + 2) (c + c^2 - d
(1 + d))$; Eq. (2g) -- the factor $(a + b - c - d + 1)$ is to be
replaced by $(a + b - c - d - 1)$; Eq. (9g) -- the term $+4$ is
missing in the square bracket, an overall factor of $\frac{1}{2}$
is missing; Eqs. (10g), (11g), (12g), (13g) -- the factor
$\sqrtpw{(2 c + 1) (2 d + 1)}$ is missing; Eq. (14h) -- a huge
expression turns into a simple one when the factor $(a - b - c - d
+ 1)$ is replaced by the correct value $(a - b + c + d + 1)$;
expression for X (table 9) -- a huge expression turns into a
simple one when the factor $(4 a b + 2 a + 4 b^2 + 3 b - 1)$ is
replaced by the correct value $(4 a b + 2 a + 2 b^2 + 3 b - 1)$;
Eqs. (1h) to (14h) -- generic expression for the coefficients is
wrong.

Substantial number of errors in Wong's paper results in its practical
unusability. Owing to this fact, we performed a detailed analysis of the
polynomial expressions and the orthogonality relations satisfied by the CG
coefficients. As a result, a self-contained list of the correct expressions of
all reduced $SO(5)$ CG coefficients involving the $14$-dimensional irreducible
representation is obtained and presented below. These formulas were
furthermore checked, using Wolfram Mathematica, in two ways. Firstly, an
Mathematica algorithm was devised that produces numerical values for the
$SO(5)$ CG coefficients. Thus determined values were checked at many points to
coincide with the values evaluated by making use of analytic
expressions. Secondly, analytic expressions for the CG coefficients were
rechecked to satisfy required orthogonality relations.

We use the same $SO(5)$ basis as Wong and Hecht \cite{R18}. However, the
notation used here differs from Wong' one; it is a natural, customary notation
of mathematical physics that reflect the "physical meaning" of relevant
variables and is designed for straightforward physical applications. In
particular, the $SO(5)$ representation labels are here denoted by $(\bar{j}_1,
\bar{j}_2)$ instead of $(a,b)$, while the irreducible representations of the
two commuting $SO(3) \supset SO(2)$ subgroup chains are here denoted by $(j_1,
m_1)$ and $(j_2, m_2)$ instead of $(c, c_m)$ and $(d, d_m)$.

\section{Reduced $SO(5)$ coefficients}

An $SO(5)$ CG coefficient can be factorized as a product of two
$SO(3)$ CG coefficients and one reduced
$SO(5)$ CG coefficient: %
\be \left(\begin{array}{@{}c@{\;}c@{}|c@{\;}c@{\;\;}c@{\;}c@{}}
 \bar{j}_1& \bar{j}_2 & \bar{j'}_1 & \bar{j'}_2 & \bar{j''}_1 & \bar{j''}_2 \\
 j_1 & j_2 & j'_1 & j'_2 & j''_1 & j''_2 \\
m_1 & m_2 & m'_1 & m'_2 & m''_1 & m''_2
\end{array}\right) =
 \left(\begin{array}{@{}c@{\;}c@{}||c@{\;}c@{\;\;}c@{\;}c@{}}
 \bar{j}_1& \bar{j}_2 & \bar{j'}_1 & \bar{j'}_2 & \bar{j''}_1 & \bar{j''}_2 \\
 j_1 & j_2 & j'_1 & j'_2 & j''_1 & j''_2
\end{array}\right)
\left(\begin{array}{@{}c|c@{\;}c@{}}
j_1 & j'_1 & j''_1 \\
m_1 & m'_1 & m''_1
\end{array}\right)
\left(\begin{array}{@{}c|c@{\;}c@{}}
j_2 & j'_2 & j''_2 \\
m_2 & m'_2 & m''_2
\end{array}\right).
\ee %
Owing to the fact that the $SO(3)$ CG coefficients are well known, only the
reduced $SO(5)$ coefficients will be listed.

Direct product of the representation $(\bar{j}_1, \bar{j}_2)$
with the $14$-dimensional representation $(\bar 1, \bar 1)$,
reduces, in general, into the following representations:%
\bed (\bar{j}_1, \bar{j}_2) \otimes (\bar 1, \bar 1) =
(\bar{j}_1+1, \bar{j}_2+1) \oplus (\bar{j}_1, \bar{j}_2+1)
\oplus(\bar{j}_1-1, \bar{j}_2+1) \oplus(\bar{j}_1+1, \bar{j}_2)
\oplus(\bar{j}_1-1, \bar{j}_2) \oplus(\bar{j}_1+1, \bar{j}_2-1)
\oplus(\bar{j}_1, \bar{j}_2-1) \oplus(\bar{j}_1-1, \bar{j}_2-1)
\oplus(\bar{j}_1+\frac 12, \bar{j}_2+\frac 12)
\oplus(\bar{j}_1-\frac 12, \bar{j}_2+\frac 12)
\oplus(\bar{j}_1+\frac 12, \bar{j}_2-\frac 12)
\oplus(\bar{j}_1-\frac 12, \bar{j}_2+\frac 12) \oplus 2(\bar{j}_1,
\bar{j}_2). \eed

\vspace{1cm} The CG coefficients $(\bar{j}_1, \bar{j}_2) \otimes
(\bar 1, \bar 1) \rightarrow (\bar{j}_1+1, \bar{j}_2+1)$ are:

\bed \text{\Nac}\left(\bar{j}_1,\bar{j}_2\right) = \left(\left(2
\bar{j}_1+2\right) \left(2 \bar{j}_1+3\right)
\left(\bar{j}_1+\bar{j}_2+2\right)
\left(\bar{j}_1+\bar{j}_2+3\right) \left(2 \bar{j}_2+1\right)
\left(2 \bar{j}_2+2\right) \left(2 \bar{j}_1+2 \bar{j}_2+3\right)
\left(2 \bar{j}_1+2 \bar{j}_2+5\right)\right)^{-\frac 12}
\eed,\bed \left(
\begin{array}{@{}c@{\;}c@{}||c@{\;}c@{\;\;}c@{\;}c@{}}
 \bar{j}_1+1 & \bar{j}_2+1 & \bar{j}_1 & \bar{j}_2 & 1 & 1 \\
 j_1+1 & j_2+1 & j_1 & j_2 & 1 & 1
\end{array}
\right)=\zfrac{\text{\Nac}\left(\bar{j}_1,\bar{j}_2\right) \
\sqrtpw{\left(j_1+j_2+\bar{j}_1-\bar{j}_2+2\right) \
\left(j_1+j_2+\bar{j}_1-\bar{j}_2+3\right) \
\left(j_1+j_2-\bar{j}_1+\bar{j}_2+1\right) \
\left(j_1+j_2-\bar{j}_1+\bar{j}_2+2\right) \
\left(j_1-j_2+\bar{j}_1+\bar{j}_2+2\right) \
\left(j_1-j_2+\bar{j}_1+\bar{j}_2+3\right) \
\left(-j_1+j_2+\bar{j}_1+\bar{j}_2+2\right)
\left(-j_1+j_2+\bar{j}_1+\ \bar{j}_2+3\right)
\left(j_1+j_2+\bar{j}_1+\bar{j}_2+3\right) \
\left(j_1+j_2+\bar{j}_1+\bar{j}_2+4\right) \
\left(j_1+j_2+\bar{j}_1+\bar{j}_2+5\right) \
\left(j_1+j_2+\bar{j}_1+\bar{j}_2+6\right)}}{4
\sqrtpw{\left(j_1+1\right) \left(2 j_1+3\right) \left(j_2+1\right)
\left(2 j_2+3\right)}}
\eed,\bed %
\left(
\begin{array}{@{}c@{\;}c@{}||c@{\;}c@{\;\;}c@{\;}c@{}}
 \bar{j}_1+1 & \bar{j}_2+1 & \bar{j}_1 & \bar{j}_2 & 1 & 1 \\
 j_1-1 & j_2-1 & j_1 & j_2 & 1 & 1
\end{array}
\right)=\zfrac{\text{\Nac}\left(\bar{j}_1,\bar{j}_2\right) \
\sqrtpw{\left(j_1+j_2+\bar{j}_1-\bar{j}_2\right) \
\left(j_1+j_2+\bar{j}_1-\bar{j}_2+1\right) \
\left(j_1+j_2-\bar{j}_1+\bar{j}_2-1\right) \
\left(j_1+j_2-\bar{j}_1+\bar{j}_2\right) \
\left(-j_1-j_2+\bar{j}_1+\bar{j}_2+1\right)
\left(-j_1-j_2+\bar{j}_1+\ \bar{j}_2+2\right)
\left(-j_1-j_2+\bar{j}_1+\bar{j}_2+3\right) \
\left(-j_1-j_2+\bar{j}_1+\bar{j}_2+4\right) \
\left(j_1-j_2+\bar{j}_1+\bar{j}_2+2\right) \
\left(j_1-j_2+\bar{j}_1+\bar{j}_2+3\right) \
\left(-j_1+j_2+\bar{j}_1+\bar{j}_2+2\right)
\left(-j_1+j_2+\bar{j}_1+\ \bar{j}_2+3\right)}}{4 \sqrtpw{j_1
\left(2 j_1-1\right) j_2 \left(2 \ j_2-1\right)}}
\eed,\bed %
\left(
\begin{array}{@{}c@{\;}c@{}||c@{\;}c@{\;\;}c@{\;}c@{}}
 \bar{j}_1+1 & \bar{j}_2+1 & \bar{j}_1 & \bar{j}_2 & 1 & 1 \\
 j_1-1 & j_2+1 & j_1 & j_2 & 1 & 1
\end{array}
\right)=\zfrac{\text{\Nac}\left(\bar{j}_1,\bar{j}_2\right) \
\sqrtpw{\left(j_1-j_2+\bar{j}_1-\bar{j}_2-1\right) \
\left(j_1-j_2+\bar{j}_1-\bar{j}_2\right) \
\left(-j_1+j_2+\bar{j}_1-\bar{j}_2+1\right)
\left(-j_1+j_2+\bar{j}_1-\ \bar{j}_2+2\right)
\left(-j_1-j_2+\bar{j}_1+\bar{j}_2+1\right) \
\left(-j_1-j_2+\bar{j}_1+\bar{j}_2+2\right)
\left(-j_1+j_2+\bar{j}_1+\ \bar{j}_2+2\right)
\left(-j_1+j_2+\bar{j}_1+\bar{j}_2+3\right) \
\left(-j_1+j_2+\bar{j}_1+\bar{j}_2+4\right)
\left(-j_1+j_2+\bar{j}_1+\ \bar{j}_2+5\right)
\left(j_1+j_2+\bar{j}_1+\bar{j}_2+3\right) \
\left(j_1+j_2+\bar{j}_1+\bar{j}_2+4\right)}}{4 \sqrtpw{j_1 \left(2
\ j_1-1\right) \left(2 j_2^2+5 j_2+3\right)}}
\eed,\bed %
\left(
\begin{array}{@{}c@{\;}c@{}||c@{\;}c@{\;\;}c@{\;}c@{}}
 \bar{j}_1+1 & \bar{j}_2+1 & \bar{j}_1 & \bar{j}_2 & 1 & 1 \\
 j_1+1 & j_2-1 & j_1 & j_2 & 1 & 1
\end{array}
\right)=\zfrac{\text{\Nac}\left(\bar{j}_1,\bar{j}_2\right) \
\sqrtpw{\left(j_1-j_2+\bar{j}_1-\bar{j}_2+1\right) \
\left(j_1-j_2+\bar{j}_1-\bar{j}_2+2\right) \
\left(-j_1+j_2+\bar{j}_1-\bar{j}_2-1\right)
\left(-j_1+j_2+\bar{j}_1-\ \bar{j}_2\right)
\left(-j_1-j_2+\bar{j}_1+\bar{j}_2+1\right) \
\left(-j_1-j_2+\bar{j}_1+\bar{j}_2+2\right) \
\left(j_1-j_2+\bar{j}_1+\bar{j}_2+2\right) \
\left(j_1-j_2+\bar{j}_1+\bar{j}_2+3\right) \
\left(j_1-j_2+\bar{j}_1+\bar{j}_2+4\right) \
\left(j_1-j_2+\bar{j}_1+\bar{j}_2+5\right) \
\left(j_1+j_2+\bar{j}_1+\bar{j}_2+3\right) \
\left(j_1+j_2+\bar{j}_1+\bar{j}_2+4\right)}}{4 \sqrtpw{\left(2
j_1^2+5 \ j_1+3\right) j_2 \left(2 j_2-1\right)}}
\eed,\bed %
\left(
\begin{array}{@{}c@{\;}c@{}||c@{\;}c@{\;\;}c@{\;}c@{}}
 \bar{j}_1+1 & \bar{j}_2+1 & \bar{j}_1 & \bar{j}_2 & 1 & 1 \\
 j_1+1 & j_2 & j_1 & j_2 & 1 & 1
\end{array}
\right)=\zfrac{\text{\Nac}\left(\bar{j}_1,\bar{j}_2\right) \
\sqrtpw{\left(j_1-j_2+\bar{j}_1-\bar{j}_2+1\right) \
\left(-j_1+j_2+\bar{j}_1-\bar{j}_2\right) \
\left(j_1+j_2+\bar{j}_1-\bar{j}_2+2\right) \
\left(j_1+j_2-\bar{j}_1+\bar{j}_2+1\right) \
\left(-j_1-j_2+\bar{j}_1+\bar{j}_2+1\right) \
\left(j_1-j_2+\bar{j}_1+\bar{j}_2+2\right) \
\left(j_1-j_2+\bar{j}_1+\bar{j}_2+3\right) \
\left(j_1-j_2+\bar{j}_1+\bar{j}_2+4\right) \
\left(-j_1+j_2+\bar{j}_1+\bar{j}_2+2\right) \
\left(j_1+j_2+\bar{j}_1+\bar{j}_2+3\right) \
\left(j_1+j_2+\bar{j}_1+\bar{j}_2+4\right) \
\left(j_1+j_2+\bar{j}_1+\bar{j}_2+5\right)}}{4
\sqrtpw{\left(j_1+1\right) \left(2 j_1+3\right) j_2
\left(j_2+1\right)}}
\eed,\bed %
\left(
\begin{array}{@{}c@{\;}c@{}||c@{\;}c@{\;\;}c@{\;}c@{}}
 \bar{j}_1+1 & \bar{j}_2+1 & \bar{j}_1 & \bar{j}_2 & 1 & 1 \\
 j_1-1 & j_2 & j_1 & j_2 & 1 & 1
\end{array}
\right)=-\zfrac{\text{\Nac}\left(\bar{j}_1,\bar{j}_2\right) \
\sqrtpw{\left(j_1-j_2+\bar{j}_1-\bar{j}_2\right) \
\left(-j_1+j_2+\bar{j}_1-\bar{j}_2+1\right) \
\left(j_1+j_2+\bar{j}_1-\bar{j}_2+1\right) \
\left(j_1+j_2-\bar{j}_1+\bar{j}_2\right) \
\left(-j_1-j_2+\bar{j}_1+\bar{j}_2+1\right)
\left(-j_1-j_2+\bar{j}_1+\ \bar{j}_2+2\right)
\left(-j_1-j_2+\bar{j}_1+\bar{j}_2+3\right) \
\left(j_1-j_2+\bar{j}_1+\bar{j}_2+2\right) \
\left(-j_1+j_2+\bar{j}_1+\bar{j}_2+2\right)
\left(-j_1+j_2+\bar{j}_1+\ \bar{j}_2+3\right)
\left(-j_1+j_2+\bar{j}_1+\bar{j}_2+4\right) \
\left(j_1+j_2+\bar{j}_1+\bar{j}_2+3\right)}}{4 \sqrtpw{j_1 \left(2
\ j_1-1\right) j_2 \left(j_2+1\right)}}
\eed,\bed %
\left(
\begin{array}{@{}c@{\;}c@{}||c@{\;}c@{\;\;}c@{\;}c@{}}
 \bar{j}_1+1 & \bar{j}_2+1 & \bar{j}_1 & \bar{j}_2 & 1 & 1 \\
 j_1 & j_2+1 & j_1 & j_2 & 1 & 1
\end{array}
\right)=\zfrac{\text{\Nac}\left(\bar{j}_1,\bar{j}_2\right) \
\sqrtpw{\left(j_1-j_2+\bar{j}_1-\bar{j}_2\right) \
\left(-j_1+j_2+\bar{j}_1-\bar{j}_2+1\right) \
\left(j_1+j_2+\bar{j}_1-\bar{j}_2+2\right) \
\left(j_1+j_2-\bar{j}_1+\bar{j}_2+1\right) \
\left(-j_1-j_2+\bar{j}_1+\bar{j}_2+1\right) \
\left(j_1-j_2+\bar{j}_1+\bar{j}_2+2\right) \
\left(-j_1+j_2+\bar{j}_1+\bar{j}_2+2\right)
\left(-j_1+j_2+\bar{j}_1+\ \bar{j}_2+3\right)
\left(-j_1+j_2+\bar{j}_1+\bar{j}_2+4\right) \
\left(j_1+j_2+\bar{j}_1+\bar{j}_2+3\right) \
\left(j_1+j_2+\bar{j}_1+\bar{j}_2+4\right) \
\left(j_1+j_2+\bar{j}_1+\bar{j}_2+5\right)}}{4 \sqrtpw{j_1
\left(j_1+1\right) \left(j_2+1\right) \left(2 j_2+3\right)}}
\eed,\bed %
\left(
\begin{array}{@{}c@{\;}c@{}||c@{\;}c@{\;\;}c@{\;}c@{}}
 \bar{j}_1+1 & \bar{j}_2+1 & \bar{j}_1 & \bar{j}_2 & 1 & 1 \\
 j_1 & j_2-1 & j_1 & j_2 & 1 & 1
\end{array}
\right)=-\zfrac{\text{\Nac}\left(\bar{j}_1,\bar{j}_2\right) \
\sqrtpw{\left(j_1-j_2+\bar{j}_1-\bar{j}_2+1\right) \
\left(-j_1+j_2+\bar{j}_1-\bar{j}_2\right) \
\left(j_1+j_2+\bar{j}_1-\bar{j}_2+1\right) \
\left(j_1+j_2-\bar{j}_1+\bar{j}_2\right) \
\left(-j_1-j_2+\bar{j}_1+\bar{j}_2+1\right)
\left(-j_1-j_2+\bar{j}_1+\ \bar{j}_2+2\right)
\left(-j_1-j_2+\bar{j}_1+\bar{j}_2+3\right) \
\left(j_1-j_2+\bar{j}_1+\bar{j}_2+2\right) \
\left(j_1-j_2+\bar{j}_1+\bar{j}_2+3\right) \
\left(j_1-j_2+\bar{j}_1+\bar{j}_2+4\right) \
\left(-j_1+j_2+\bar{j}_1+\bar{j}_2+2\right) \
\left(j_1+j_2+\bar{j}_1+\bar{j}_2+3\right)}}{4 \sqrtpw{j_1
\left(j_1+1\right) j_2 \left(2 j_2-1\right)}}
\eed,\bed %
\left(
\begin{array}{@{}c@{\;}c@{}||c@{\;}c@{\;\;}c@{\;}c@{}}
 \bar{j}_1+1 & \bar{j}_2+1 & \bar{j}_1 & \bar{j}_2 & 1 & 1 \\
 j_1 & j_2 & j_1 & j_2 & 1 & 1
\end{array}
\right)=-\zfrac{\text{\Nac}\left(\bar{j}_1,\bar{j}_2\right) \
\sqrtpw{\left(-j_1-j_2+\bar{j}_1+\bar{j}_2+1\right) \
\left(-j_1-j_2+\bar{j}_1+\bar{j}_2+2\right) \
\left(j_1-j_2+\bar{j}_1+\bar{j}_2+2\right) \
\left(j_1-j_2+\bar{j}_1+\bar{j}_2+3\right) \
\left(-j_1+j_2+\bar{j}_1+\bar{j}_2+2\right)
\left(-j_1+j_2+\bar{j}_1+\ \bar{j}_2+3\right)
\left(j_1+j_2+\bar{j}_1+\bar{j}_2+3\right) \
\left(j_1+j_2+\bar{j}_1+\bar{j}_2+4\right)} \
\left(j_1^2+j_1+j_2^2-\bar{j}_1^2-\bar{j}_2^2+j_2-\bar{j}_1+2 \
\bar{j}_1 \bar{j}_2+\bar{j}_2\right)}{4 \sqrtpw{j_1
\left(j_1+1\right) \ j_2 \left(j_2+1\right)}}
\eed,\bed %
\left(
\begin{array}{@{}c@{\;}c@{}||c@{\;}c@{\;\;}c@{\;}c@{}}
 \bar{j}_1+1 & \bar{j}_2+1 & \bar{j}_1 & \bar{j}_2 & 1 & 1 \\
 j_1+\frac{1}{2} & j_2+\frac{1}{2} & j_1 & j_2 & \frac{1}{2} & \
\frac{1}{2}
\end{array}
\right)=\zfrac{\text{\Nac}\left(\bar{j}_1,\bar{j}_2\right) \
\sqrtpw{\left(j_1+j_2+\bar{j}_1-\bar{j}_2+2\right) \
\left(j_1+j_2-\bar{j}_1+\bar{j}_2+1\right) \
\left(-j_1-j_2+\bar{j}_1+\bar{j}_2+1\right) \
\left(j_1-j_2+\bar{j}_1+\bar{j}_2+2\right) \
\left(j_1-j_2+\bar{j}_1+\bar{j}_2+3\right) \
\left(-j_1+j_2+\bar{j}_1+\bar{j}_2+2\right)
\left(-j_1+j_2+\bar{j}_1+\ \bar{j}_2+3\right)
\left(j_1+j_2+\bar{j}_1+\bar{j}_2+3\right) \
\left(j_1+j_2+\bar{j}_1+\bar{j}_2+4\right) \
\left(j_1+j_2+\bar{j}_1+\bar{j}_2+5\right)}}{2
\sqrtpw{\left(j_1+1\right) \left(j_2+1\right)}}
\eed,\bed %
\left(
\begin{array}{@{}c@{\;}c@{}||c@{\;}c@{\;\;}c@{\;}c@{}}
 \bar{j}_1+1 & \bar{j}_2+1 & \bar{j}_1 & \bar{j}_2 & 1 & 1 \\
 j_1-\frac{1}{2} & j_2-\frac{1}{2} & j_1 & j_2 & \frac{1}{2} & \
\frac{1}{2}
\end{array}
\right)=-\zfrac{\text{\Nac}\left(\bar{j}_1,\bar{j}_2\right) \
\sqrtpw{\left(j_1+j_2+\bar{j}_1-\bar{j}_2+1\right) \
\left(j_1+j_2-\bar{j}_1+\bar{j}_2\right) \
\left(-j_1-j_2+\bar{j}_1+\bar{j}_2+1\right)
\left(-j_1-j_2+\bar{j}_1+\ \bar{j}_2+2\right)
\left(-j_1-j_2+\bar{j}_1+\bar{j}_2+3\right) \
\left(j_1-j_2+\bar{j}_1+\bar{j}_2+2\right) \
\left(j_1-j_2+\bar{j}_1+\bar{j}_2+3\right) \
\left(-j_1+j_2+\bar{j}_1+\bar{j}_2+2\right)
\left(-j_1+j_2+\bar{j}_1+\ \bar{j}_2+3\right)
\left(j_1+j_2+\bar{j}_1+\bar{j}_2+3\right)}}{2 \ \sqrtpw{j_1 j_2}}
\eed,\bed %
\left(
\begin{array}{@{}c@{\;}c@{}||c@{\;}c@{\;\;}c@{\;}c@{}}
 \bar{j}_1+1 & \bar{j}_2+1 & \bar{j}_1 & \bar{j}_2 & 1 & 1 \\
 j_1+\frac{1}{2} & j_2-\frac{1}{2} & j_1 & j_2 & \frac{1}{2} & \
\frac{1}{2}
\end{array}
\right)=\zfrac{\text{\Nac}\left(\bar{j}_1,\bar{j}_2\right) \
\sqrtpw{\left(j_1-j_2+\bar{j}_1-\bar{j}_2+1\right) \
\left(-j_1+j_2+\bar{j}_1-\bar{j}_2\right) \
\left(-j_1-j_2+\bar{j}_1+\bar{j}_2+1\right)
\left(-j_1-j_2+\bar{j}_1+\ \bar{j}_2+2\right)
\left(j_1-j_2+\bar{j}_1+\bar{j}_2+2\right) \
\left(j_1-j_2+\bar{j}_1+\bar{j}_2+3\right) \
\left(j_1-j_2+\bar{j}_1+\bar{j}_2+4\right) \
\left(-j_1+j_2+\bar{j}_1+\bar{j}_2+2\right) \
\left(j_1+j_2+\bar{j}_1+\bar{j}_2+3\right) \
\left(j_1+j_2+\bar{j}_1+\bar{j}_2+4\right)}}{2
\sqrtpw{\left(j_1+1\right) j_2}}
\eed,\bed %
\left(
\begin{array}{@{}c@{\;}c@{}||c@{\;}c@{\;\;}c@{\;}c@{}}
 \bar{j}_1+1 & \bar{j}_2+1 & \bar{j}_1 & \bar{j}_2 & 1 & 1 \\
 j_1-\frac{1}{2} & j_2+\frac{1}{2} & j_1 & j_2 & \frac{1}{2} & \
\frac{1}{2}
\end{array}
\right)=\zfrac{\text{\Nac}\left(\bar{j}_1,\bar{j}_2\right) \
\sqrtpw{\left(j_1-j_2+\bar{j}_1-\bar{j}_2\right) \
\left(-j_1+j_2+\bar{j}_1-\bar{j}_2+1\right)
\left(-j_1-j_2+\bar{j}_1+\ \bar{j}_2+1\right)
\left(-j_1-j_2+\bar{j}_1+\bar{j}_2+2\right) \
\left(j_1-j_2+\bar{j}_1+\bar{j}_2+2\right) \
\left(-j_1+j_2+\bar{j}_1+\bar{j}_2+2\right)
\left(-j_1+j_2+\bar{j}_1+\ \bar{j}_2+3\right)
\left(-j_1+j_2+\bar{j}_1+\bar{j}_2+4\right) \
\left(j_1+j_2+\bar{j}_1+\bar{j}_2+3\right) \
\left(j_1+j_2+\bar{j}_1+\bar{j}_2+4\right)}}{2 \sqrtpw{j_1
\left(j_2+1\right)}}
\eed,\bed %
\left(
\begin{array}{@{}c@{\;}c@{}||c@{\;}c@{\;\;}c@{\;}c@{}}
 \bar{j}_1+1 & \bar{j}_2+1 & \bar{j}_1 & \bar{j}_2 & 1 & 1 \\
 j_1 & j_2 & j_1 & j_2 & 0 & 0
\end{array}
\right)=\frac{1}{2}\sqrtpw{5}
\text{\Nac}\left(\bar{j}_1,\bar{j}_2\right) \
\sqrtpw{\left(-j_1-j_2+\bar{j}_1+\bar{j}_2+1\right) \
\left(-j_1-j_2+\bar{j}_1+\bar{j}_2+2\right) \
\left(j_1-j_2+\bar{j}_1+\bar{j}_2+2\right) \
\left(j_1-j_2+\bar{j}_1+\bar{j}_2+3\right) \
\left(-j_1+j_2+\bar{j}_1+\bar{j}_2+2\right)
\left(-j_1+j_2+\bar{j}_1+\ \bar{j}_2+3\right)
\left(j_1+j_2+\bar{j}_1+\bar{j}_2+3\right) \
\left(j_1+j_2+\bar{j}_1+\bar{j}_2+4\right)}
\eed %
.

\vspace{1cm}
The CG coefficients $(\bar{j}_1, \bar{j}_2) \otimes
(\bar 1, \bar 1) \rightarrow (\bar{j}_1+1, \bar{j}_2)$ are:

\bed \text{\Nbc}\left(\bar{j}_1,\bar{j}_2\right) = \left(\left(2
\bar{j}_1+2\right) \left(2 \bar{j}_1+3\right) \left(2 \bar{j}_1-2
\bar{j}_2+1\right) \left(\bar{j}_1-\bar{j}_2+1\right)
   \bar{j}_2 \left(\bar{j}_1+\bar{j}_2+2\right) \left(2 \bar{j}_2+2\right)
   \left(2 \bar{j}_1+2 \bar{j}_2+3\right)\right)^{-\frac 12} \eed

\bed%
\left(
\begin{array}{@{}c@{\;}c@{}||c@{\;}c@{\;\;}c@{\;}c@{}}
 \bar{j}_1+1 & \bar{j}_2 & \bar{j}_1 & \bar{j}_2 & 1 & 1 \\
 j_1+1 & j_2+1 & j_1 & j_2 & 1 & 1
\end{array}
\right)=-\zfrac{\text{\Nbc}\left(\bar{j}_1,\bar{j}_2\right) \
\sqrtpw{\left(j_1-j_2+\bar{j}_1-\bar{j}_2+1\right) \
\left(-j_1+j_2+\bar{j}_1-\bar{j}_2+1\right) \
\left(j_1+j_2+\bar{j}_1-\bar{j}_2+2\right) \
\left(j_1+j_2+\bar{j}_1-\bar{j}_2+3\right) \
\left(j_1+j_2+\bar{j}_1-\bar{j}_2+4\right) \
\left(j_1+j_2-\bar{j}_1+\bar{j}_2+1\right) \
\left(-j_1-j_2+\bar{j}_1+\bar{j}_2\right) \
\left(j_1-j_2+\bar{j}_1+\bar{j}_2+2\right) \
\left(-j_1+j_2+\bar{j}_1+\bar{j}_2+2\right) \
\left(j_1+j_2+\bar{j}_1+\bar{j}_2+3\right) \
\left(j_1+j_2+\bar{j}_1+\bar{j}_2+4\right) \
\left(j_1+j_2+\bar{j}_1+\bar{j}_2+5\right)}}{4
\sqrtpw{\left(j_1+1\right) \left(2 j_1+3\right) \left(j_2+1\right)
\left(2 j_2+3\right)}}
\eed,\bed%
\left(
\begin{array}{@{}c@{\;}c@{}||c@{\;}c@{\;\;}c@{\;}c@{}}
 \bar{j}_1+1 & \bar{j}_2 & \bar{j}_1 & \bar{j}_2 & 1 & 1 \\
 j_1-1 & j_2-1 & j_1 & j_2 & 1 & 1
\end{array}
\right)=\zfrac{\text{\Nbc}\left(\bar{j}_1,\bar{j}_2\right) \
\sqrtpw{\left(j_1-j_2+\bar{j}_1-\bar{j}_2+1\right) \
\left(-j_1+j_2+\bar{j}_1-\bar{j}_2+1\right) \
\left(j_1+j_2+\bar{j}_1-\bar{j}_2+1\right) \
\left(j_1+j_2-\bar{j}_1+\bar{j}_2-2\right) \
\left(j_1+j_2-\bar{j}_1+\bar{j}_2-1\right) \
\left(j_1+j_2-\bar{j}_1+\bar{j}_2\right) \
\left(-j_1-j_2+\bar{j}_1+\bar{j}_2+1\right)
\left(-j_1-j_2+\bar{j}_1+\ \bar{j}_2+2\right)
\left(-j_1-j_2+\bar{j}_1+\bar{j}_2+3\right) \
\left(j_1-j_2+\bar{j}_1+\bar{j}_2+2\right) \
\left(-j_1+j_2+\bar{j}_1+\bar{j}_2+2\right) \
\left(j_1+j_2+\bar{j}_1+\bar{j}_2+2\right)}}{4 \sqrtpw{j_1 \left(2
\ j_1-1\right) j_2 \left(2 j_2-1\right)}}
\eed,\bed%
\left(
\begin{array}{@{}c@{\;}c@{}||c@{\;}c@{\;\;}c@{\;}c@{}}
 \bar{j}_1+1 & \bar{j}_2 & \bar{j}_1 & \bar{j}_2 & 1 & 1 \\
 j_1-1 & j_2+1 & j_1 & j_2 & 1 & 1
\end{array}
\right)=\zfrac{\text{\Nbc}\left(\bar{j}_1,\bar{j}_2\right) \
\sqrtpw{\left(j_1-j_2+\bar{j}_1-\bar{j}_2\right) \
\left(-j_1+j_2+\bar{j}_1-\bar{j}_2+1\right)
\left(-j_1+j_2+\bar{j}_1-\ \bar{j}_2+2\right)
\left(-j_1+j_2+\bar{j}_1-\bar{j}_2+3\right) \
\left(j_1+j_2+\bar{j}_1-\bar{j}_2+2\right) \
\left(j_1+j_2-\bar{j}_1+\bar{j}_2\right) \
\left(-j_1-j_2+\bar{j}_1+\bar{j}_2+1\right) \
\left(j_1-j_2+\bar{j}_1+\bar{j}_2+1\right) \
\left(-j_1+j_2+\bar{j}_1+\bar{j}_2+2\right)
\left(-j_1+j_2+\bar{j}_1+\ \bar{j}_2+3\right)
\left(-j_1+j_2+\bar{j}_1+\bar{j}_2+4\right) \
\left(j_1+j_2+\bar{j}_1+\bar{j}_2+3\right)}}{4 \sqrtpw{j_1 \left(2
\ j_1-1\right) \left(2 j_2^2+5 j_2+3\right)}}
\eed,\bed%
\left(
\begin{array}{@{}c@{\;}c@{}||c@{\;}c@{\;\;}c@{\;}c@{}}
 \bar{j}_1+1 & \bar{j}_2 & \bar{j}_1 & \bar{j}_2 & 1 & 1 \\
 j_1+1 & j_2-1 & j_1 & j_2 & 1 & 1
\end{array}
\right)=\zfrac{\text{\Nbc}\left(\bar{j}_1,\bar{j}_2\right) \
\sqrtpw{\left(j_1-j_2+\bar{j}_1-\bar{j}_2+1\right) \
\left(j_1-j_2+\bar{j}_1-\bar{j}_2+2\right) \
\left(j_1-j_2+\bar{j}_1-\bar{j}_2+3\right) \
\left(-j_1+j_2+\bar{j}_1-\bar{j}_2\right) \
\left(j_1+j_2+\bar{j}_1-\bar{j}_2+2\right) \
\left(j_1+j_2-\bar{j}_1+\bar{j}_2\right) \
\left(-j_1-j_2+\bar{j}_1+\bar{j}_2+1\right) \
\left(j_1-j_2+\bar{j}_1+\bar{j}_2+2\right) \
\left(j_1-j_2+\bar{j}_1+\bar{j}_2+3\right) \
\left(j_1-j_2+\bar{j}_1+\bar{j}_2+4\right) \
\left(-j_1+j_2+\bar{j}_1+\bar{j}_2+1\right) \
\left(j_1+j_2+\bar{j}_1+\bar{j}_2+3\right)}}{4 \sqrtpw{\left(2
j_1^2+5 \ j_1+3\right) j_2 \left(2 j_2-1\right)}}
\eed,\bed%
\left(
\begin{array}{@{}c@{\;}c@{}||c@{\;}c@{\;\;}c@{\;}c@{}}
 \bar{j}_1+1 & \bar{j}_2 & \bar{j}_1 & \bar{j}_2 & 1 & 1 \\
 j_1+1 & j_2 & j_1 & j_2 & 1 & 1
\end{array}
\right)=\zfrac{\text{\Nbc}\left(\bar{j}_1,\bar{j}_2\right) \
\sqrtpw{\left(j_1-j_2+\bar{j}_1-\bar{j}_2+1\right) \
\left(j_1-j_2+\bar{j}_1-\bar{j}_2+2\right) \
\left(j_1+j_2+\bar{j}_1-\bar{j}_2+2\right) \
\left(j_1+j_2+\bar{j}_1-\bar{j}_2+3\right) \
\left(j_1-j_2+\bar{j}_1+\bar{j}_2+2\right) \
\left(j_1-j_2+\bar{j}_1+\bar{j}_2+3\right) \
\left(j_1+j_2+\bar{j}_1+\bar{j}_2+3\right) \
\left(j_1+j_2+\bar{j}_1+\bar{j}_2+4\right)} \left(-j_1^2+\left(2 \
\bar{j}_1+1\right) \
j_1+j_2^2-\bar{j}_1^2+\bar{j}_2^2+j_2-\bar{j}_1+\bar{j}_2\right)}{4
\ \sqrtpw{\left(2 j_1^2+5 j_1+3\right) j_2 \left(j_2+1\right)}}
\eed,\bed%
\left(
\begin{array}{@{}c@{\;}c@{}||c@{\;}c@{\;\;}c@{\;}c@{}}
 \bar{j}_1+1 & \bar{j}_2 & \bar{j}_1 & \bar{j}_2 & 1 & 1 \\
 j_1-1 & j_2 & j_1 & j_2 & 1 & 1
\end{array}
\right)=-\zfrac{\text{\Nbc}\left(\bar{j}_1,\bar{j}_2\right) \
\sqrtpw{\left(-j_1+j_2+\bar{j}_1-\bar{j}_2+1\right) \
\left(-j_1+j_2+\bar{j}_1-\bar{j}_2+2\right) \
\left(j_1+j_2-\bar{j}_1+\bar{j}_2-1\right) \
\left(j_1+j_2-\bar{j}_1+\bar{j}_2\right) \
\left(-j_1-j_2+\bar{j}_1+\bar{j}_2+1\right)
\left(-j_1-j_2+\bar{j}_1+\ \bar{j}_2+2\right)
\left(-j_1+j_2+\bar{j}_1+\bar{j}_2+2\right) \
\left(-j_1+j_2+\bar{j}_1+\bar{j}_2+3\right)} \left(2 j_2 \
\bar{j}_2+\left(j_1-j_2+\bar{j}_1-\bar{j}_2+1\right) \
\left(j_1+j_2+\bar{j}_1+\bar{j}_2+2\right)\right)}{4 \sqrtpw{j_1 \
\left(2 j_1-1\right) j_2 \left(j_2+1\right)}}
\eed,\bed%
\left(
\begin{array}{@{}c@{\;}c@{}||c@{\;}c@{\;\;}c@{\;}c@{}}
 \bar{j}_1+1 & \bar{j}_2 & \bar{j}_1 & \bar{j}_2 & 1 & 1 \\
 j_1 & j_2+1 & j_1 & j_2 & 1 & 1
\end{array}
\right)=\zfrac{\text{\Nbc}\left(\bar{j}_1,\bar{j}_2\right) \
\sqrtpw{\left(-j_1+j_2+\bar{j}_1-\bar{j}_2+1\right) \
\left(-j_1+j_2+\bar{j}_1-\bar{j}_2+2\right) \
\left(j_1+j_2+\bar{j}_1-\bar{j}_2+2\right) \
\left(j_1+j_2+\bar{j}_1-\bar{j}_2+3\right) \
\left(-j_1+j_2+\bar{j}_1+\bar{j}_2+2\right)
\left(-j_1+j_2+\bar{j}_1+\ \bar{j}_2+3\right)
\left(j_1+j_2+\bar{j}_1+\bar{j}_2+3\right) \
\left(j_1+j_2+\bar{j}_1+\bar{j}_2+4\right)} \
\left(j_1^2+j_1-j_2^2-\bar{j}_1^2+\bar{j}_2^2-\bar{j}_1+j_2
\left(2 \ \bar{j}_1+1\right)+\bar{j}_2\right)}{4 \sqrtpw{j_1
\left(j_1+1\right) \ \left(2 j_2^2+5 j_2+3\right)}}
\eed,\bed%
\left(
\begin{array}{@{}c@{\;}c@{}||c@{\;}c@{\;\;}c@{\;}c@{}}
 \bar{j}_1+1 & \bar{j}_2 & \bar{j}_1 & \bar{j}_2 & 1 & 1 \\
 j_1 & j_2-1 & j_1 & j_2 & 1 & 1
\end{array}
\right)=\zfrac{\text{\Nbc}\left(\bar{j}_1,\bar{j}_2\right) \
\sqrtpw{\left(j_1-j_2+\bar{j}_1-\bar{j}_2+1\right) \
\left(j_1-j_2+\bar{j}_1-\bar{j}_2+2\right) \
\left(j_1+j_2-\bar{j}_1+\bar{j}_2-1\right) \
\left(j_1+j_2-\bar{j}_1+\bar{j}_2\right) \
\left(-j_1-j_2+\bar{j}_1+\bar{j}_2+1\right)
\left(-j_1-j_2+\bar{j}_1+\ \bar{j}_2+2\right)
\left(j_1-j_2+\bar{j}_1+\bar{j}_2+2\right) \
\left(j_1-j_2+\bar{j}_1+\bar{j}_2+3\right)} \
\left(j_1^2+j_1-j_2^2-\bar{j}_1^2+\bar{j}_2^2-3 \bar{j}_1-j_2
\left(2 \ \bar{j}_1+3\right)+\bar{j}_2-2\right)}{4 \sqrtpw{j_1
\left(j_1+1\right) \ j_2 \left(2 j_2-1\right)}}
\eed,\bed%
\left(
\begin{array}{@{}c@{\;}c@{}||c@{\;}c@{\;\;}c@{\;}c@{}}
 \bar{j}_1+1 & \bar{j}_2 & \bar{j}_1 & \bar{j}_2 & 1 & 1 \\
 j_1 & j_2 & j_1 & j_2 & 1 & 1
\end{array}
\right)=-\zfrac{\text{\Nbc}\left(\bar{j}_1,\bar{j}_2\right) \
\sqrtpw{\left(j_1-j_2+\bar{j}_1-\bar{j}_2+1\right) \
\left(-j_1+j_2+\bar{j}_1-\bar{j}_2+1\right) \
\left(j_1+j_2+\bar{j}_1-\bar{j}_2+2\right) \
\left(j_1+j_2-\bar{j}_1+\bar{j}_2\right) \
\left(-j_1-j_2+\bar{j}_1+\bar{j}_2+1\right) \
\left(j_1-j_2+\bar{j}_1+\bar{j}_2+2\right) \
\left(-j_1+j_2+\bar{j}_1+\bar{j}_2+2\right) \
\left(j_1+j_2+\bar{j}_1+\bar{j}_2+3\right)} \
\left(j_1^2+j_1+j_2^2-\bar{j}_1^2+\bar{j}_2^2+j_2-3 \
\bar{j}_1+\bar{j}_2-2\right)}{4 \sqrtpw{j_1 \left(j_1+1\right) j_2
\ \left(j_2+1\right)}}
\eed,\bed%
\left(
\begin{array}{@{}c@{\;}c@{}||c@{\;}c@{\;\;}c@{\;}c@{}}
 \bar{j}_1+1 & \bar{j}_2 & \bar{j}_1 & \bar{j}_2 & 1 & 1 \\
 j_1+\frac{1}{2} & j_2+\frac{1}{2} & j_1 & j_2 & \frac{1}{2} & \
\frac{1}{2}
\end{array}
\right)=\zfrac{\text{\Nbc}\left(\bar{j}_1,\bar{j}_2\right) \
\left(j_1+j_2-\bar{j}_1\right) \
\sqrtpw{\left(j_1-j_2+\bar{j}_1-\bar{j}_2+1\right) \
\left(-j_1+j_2+\bar{j}_1-\bar{j}_2+1\right) \
\left(j_1+j_2+\bar{j}_1-\bar{j}_2+2\right) \
\left(j_1+j_2+\bar{j}_1-\bar{j}_2+3\right) \
\left(j_1-j_2+\bar{j}_1+\bar{j}_2+2\right) \
\left(-j_1+j_2+\bar{j}_1+\bar{j}_2+2\right) \
\left(j_1+j_2+\bar{j}_1+\bar{j}_2+3\right) \
\left(j_1+j_2+\bar{j}_1+\bar{j}_2+4\right)}}{2
\sqrtpw{\left(j_1+1\right) \left(j_2+1\right)}}
\eed,\bed%
\left(
\begin{array}{@{}c@{\;}c@{}||c@{\;}c@{\;\;}c@{\;}c@{}}
 \bar{j}_1+1 & \bar{j}_2 & \bar{j}_1 & \bar{j}_2 & 1 & 1 \\
 j_1-\frac{1}{2} & j_2-\frac{1}{2} & j_1 & j_2 & \frac{1}{2} & \
\frac{1}{2}
\end{array}
\right)=-\zfrac{\text{\Nbc}\left(\bar{j}_1,\bar{j}_2\right) \
\left(j_1+j_2+\bar{j}_1+2\right) \
\sqrtpw{\left(j_1-j_2+\bar{j}_1-\bar{j}_2+1\right) \
\left(-j_1+j_2+\bar{j}_1-\bar{j}_2+1\right) \
\left(j_1+j_2-\bar{j}_1+\bar{j}_2-1\right) \
\left(j_1+j_2-\bar{j}_1+\bar{j}_2\right) \
\left(-j_1-j_2+\bar{j}_1+\bar{j}_2+1\right)
\left(-j_1-j_2+\bar{j}_1+\ \bar{j}_2+2\right)
\left(j_1-j_2+\bar{j}_1+\bar{j}_2+2\right) \
\left(-j_1+j_2+\bar{j}_1+\bar{j}_2+2\right)}}{2 \sqrtpw{j_1 j_2}}
\eed,\bed%
\left(
\begin{array}{@{}c@{\;}c@{}||c@{\;}c@{\;\;}c@{\;}c@{}}
 \bar{j}_1+1 & \bar{j}_2 & \bar{j}_1 & \bar{j}_2 & 1 & 1 \\
 j_1+\frac{1}{2} & j_2-\frac{1}{2} & j_1 & j_2 & \frac{1}{2} & \
\frac{1}{2}
\end{array}
\right)=\zfrac{\text{\Nbc}\left(\bar{j}_1,\bar{j}_2\right) \
\left(-j_1+j_2+\bar{j}_1+1\right) \
\sqrtpw{\left(j_1-j_2+\bar{j}_1-\bar{j}_2+1\right) \
\left(j_1-j_2+\bar{j}_1-\bar{j}_2+2\right) \
\left(j_1+j_2+\bar{j}_1-\bar{j}_2+2\right) \
\left(j_1+j_2-\bar{j}_1+\bar{j}_2\right) \
\left(-j_1-j_2+\bar{j}_1+\bar{j}_2+1\right) \
\left(j_1-j_2+\bar{j}_1+\bar{j}_2+2\right) \
\left(j_1-j_2+\bar{j}_1+\bar{j}_2+3\right) \
\left(j_1+j_2+\bar{j}_1+\bar{j}_2+3\right)}}{2
\sqrtpw{\left(j_1+1\right) j_2}}
\eed,\bed%
\left(
\begin{array}{@{}c@{\;}c@{}||c@{\;}c@{\;\;}c@{\;}c@{}}
 \bar{j}_1+1 & \bar{j}_2 & \bar{j}_1 & \bar{j}_2 & 1 & 1 \\
 j_1-\frac{1}{2} & j_2+\frac{1}{2} & j_1 & j_2 & \frac{1}{2} & \
\frac{1}{2}
\end{array}
\right)=\zfrac{\text{\Nbc}\left(\bar{j}_1,\bar{j}_2\right) \
\left(j_1-j_2+\bar{j}_1+1\right) \
\sqrtpw{\left(-j_1+j_2+\bar{j}_1-\bar{j}_2+1\right) \
\left(-j_1+j_2+\bar{j}_1-\bar{j}_2+2\right) \
\left(j_1+j_2+\bar{j}_1-\bar{j}_2+2\right) \
\left(j_1+j_2-\bar{j}_1+\bar{j}_2\right) \
\left(-j_1-j_2+\bar{j}_1+\bar{j}_2+1\right)
\left(-j_1+j_2+\bar{j}_1+\ \bar{j}_2+2\right)
\left(-j_1+j_2+\bar{j}_1+\bar{j}_2+3\right) \
\left(j_1+j_2+\bar{j}_1+\bar{j}_2+3\right)}}{2 \sqrtpw{j_1
\left(j_2+1\right)}}
\eed,\bed%
\left(
\begin{array}{@{}c@{\;}c@{}||c@{\;}c@{\;\;}c@{\;}c@{}}
 \bar{j}_1+1 & \bar{j}_2 & \bar{j}_1 & \bar{j}_2 & 1 & 1 \\
 j_1 & j_2 & j_1 & j_2 & 0 & 0
\end{array}
\right)=\frac{1}{2} \sqrtpw{5}
\text{\Nbc}\left(\bar{j}_1,\bar{j}_2\right)
\sqrtpw{\left(j_1-j_2+\bar{j}_1-\bar{j}_2+1\right) \
\left(-j_1+j_2+\bar{j}_1-\bar{j}_2+1\right) \
\left(j_1+j_2+\bar{j}_1-\bar{j}_2+2\right) \
\left(j_1+j_2-\bar{j}_1+\bar{j}_2\right) \
\left(-j_1-j_2+\bar{j}_1+\bar{j}_2+1\right) \
\left(j_1-j_2+\bar{j}_1+\bar{j}_2+2\right) \
\left(-j_1+j_2+\bar{j}_1+\bar{j}_2+2\right) \
\left(j_1+j_2+\bar{j}_1+\bar{j}_2+3\right)}
\eed%
.

\vspace{1cm} The CG coefficients $(\bar{j}_1, \bar{j}_2) \otimes
(\bar 1, \bar 1) \rightarrow (\bar{j}_1, \bar{j}_2+1)$ are:

\bed \text{\Ncc}\left(\bar{j}_1,\bar{j}_2\right) = \left(\left(2
\bar{j}_1+1\right) \left(2 \bar{j}_1+3\right) \left(2 \bar{j}_1-2
\bar{j}_2+1\right) \left(\bar{j}_1-\bar{j}_2\right)
   \left(\bar{j}_1+\bar{j}_2+2\right) \left(2 \bar{j}_2+1\right) \left(2
     \bar{j}_2+2\right) \left(2 \bar{j}_1+2 \bar{j}_2+3\right)\right)^{-\frac
   12} \eed

\bed \left(
\begin{array}{@{}c@{\;}c@{}||c@{\;}c@{\;\;}c@{\;}c@{}}
 \bar{j}_1 & \bar{j}_2+1 & \bar{j}_1 & \bar{j}_2 & 1 & 1 \\
 j_1+1 & j_2+1 & j_1 & j_2 & 1 & 1
\end{array}
\right)=\zfrac{\text{\Ncc}\left(\bar{j}_1,\bar{j}_2\right) \
\sqrtpw{\left(j_1-j_2+\bar{j}_1-\bar{j}_2\right) \
\left(-j_1+j_2+\bar{j}_1-\bar{j}_2\right) \
\left(j_1+j_2+\bar{j}_1-\bar{j}_2+2\right) \
\left(j_1+j_2-\bar{j}_1+\bar{j}_2+1\right) \
\left(j_1+j_2-\bar{j}_1+\bar{j}_2+2\right) \
\left(j_1+j_2-\bar{j}_1+\bar{j}_2+3\right) \
\left(-j_1-j_2+\bar{j}_1+\bar{j}_2\right) \
\left(j_1-j_2+\bar{j}_1+\bar{j}_2+2\right) \
\left(-j_1+j_2+\bar{j}_1+\bar{j}_2+2\right) \
\left(j_1+j_2+\bar{j}_1+\bar{j}_2+3\right) \
\left(j_1+j_2+\bar{j}_1+\bar{j}_2+4\right) \
\left(j_1+j_2+\bar{j}_1+\bar{j}_2+5\right)}}{2 \sqrtpw{2} \
\sqrtpw{\left(j_1+1\right) \left(2 j_1+3\right) \left(j_2+1\right)
\ \left(2 j_2+3\right)}}
\eed,\bed%
\left(
\begin{array}{@{}c@{\;}c@{}||c@{\;}c@{\;\;}c@{\;}c@{}}
 \bar{j}_1 & \bar{j}_2+1 & \bar{j}_1 & \bar{j}_2 & 1 & 1 \\
 j_1-1 & j_2-1 & j_1 & j_2 & 1 & 1
\end{array}
\right)=-\zfrac{\text{\Ncc}\left(\bar{j}_1,\bar{j}_2\right) \
\sqrtpw{\left(j_1-j_2+\bar{j}_1-\bar{j}_2\right) \
\left(-j_1+j_2+\bar{j}_1-\bar{j}_2\right) \
\left(j_1+j_2+\bar{j}_1-\bar{j}_2-1\right) \
\left(j_1+j_2+\bar{j}_1-\bar{j}_2\right) \
\left(j_1+j_2+\bar{j}_1-\bar{j}_2+1\right) \
\left(j_1+j_2-\bar{j}_1+\bar{j}_2\right) \
\left(-j_1-j_2+\bar{j}_1+\bar{j}_2+1\right)
\left(-j_1-j_2+\bar{j}_1+\ \bar{j}_2+2\right)
\left(-j_1-j_2+\bar{j}_1+\bar{j}_2+3\right) \
\left(j_1-j_2+\bar{j}_1+\bar{j}_2+2\right) \
\left(-j_1+j_2+\bar{j}_1+\bar{j}_2+2\right) \
\left(j_1+j_2+\bar{j}_1+\bar{j}_2+2\right)}}{2 \sqrtpw{2}
\sqrtpw{j_1 \ \left(2 j_1-1\right) j_2 \left(2 j_2-1\right)}}
\eed,\bed%
\left(
\begin{array}{@{}c@{\;}c@{}||c@{\;}c@{\;\;}c@{\;}c@{}}
 \bar{j}_1 & \bar{j}_2+1 & \bar{j}_1 & \bar{j}_2 & 1 & 1 \\
 j_1-1 & j_2+1 & j_1 & j_2 & 1 & 1
\end{array}
\right)=\zfrac{\text{\Ncc}\left(\bar{j}_1,\bar{j}_2\right) \
\sqrtpw{\left(j_1-j_2+\bar{j}_1-\bar{j}_2-2\right) \
\left(j_1-j_2+\bar{j}_1-\bar{j}_2-1\right) \
\left(j_1-j_2+\bar{j}_1-\bar{j}_2\right) \
\left(-j_1+j_2+\bar{j}_1-\bar{j}_2+1\right) \
\left(j_1+j_2+\bar{j}_1-\bar{j}_2+1\right) \
\left(j_1+j_2-\bar{j}_1+\bar{j}_2+1\right) \
\left(-j_1-j_2+\bar{j}_1+\bar{j}_2+1\right) \
\left(j_1-j_2+\bar{j}_1+\bar{j}_2+1\right) \
\left(-j_1+j_2+\bar{j}_1+\bar{j}_2+2\right)
\left(-j_1+j_2+\bar{j}_1+\ \bar{j}_2+3\right)
\left(-j_1+j_2+\bar{j}_1+\bar{j}_2+4\right) \
\left(j_1+j_2+\bar{j}_1+\bar{j}_2+3\right)}}{2 \sqrtpw{2}
\sqrtpw{j_1 \ \left(2 j_1-1\right) \left(j_2+1\right) \left(2
j_2+3\right)}}
\eed,\bed%
\left(
\begin{array}{@{}c@{\;}c@{}||c@{\;}c@{\;\;}c@{\;}c@{}}
 \bar{j}_1 & \bar{j}_2+1 & \bar{j}_1 & \bar{j}_2 & 1 & 1 \\
 j_1+1 & j_2-1 & j_1 & j_2 & 1 & 1
\end{array}
\right)=\zfrac{\text{\Ncc}\left(\bar{j}_1,\bar{j}_2\right) \
\sqrtpw{\left(j_1-j_2+\bar{j}_1-\bar{j}_2+1\right) \
\left(-j_1+j_2+\bar{j}_1-\bar{j}_2-2\right)
\left(-j_1+j_2+\bar{j}_1-\ \bar{j}_2-1\right)
\left(-j_1+j_2+\bar{j}_1-\bar{j}_2\right) \
\left(j_1+j_2+\bar{j}_1-\bar{j}_2+1\right) \
\left(j_1+j_2-\bar{j}_1+\bar{j}_2+1\right) \
\left(-j_1-j_2+\bar{j}_1+\bar{j}_2+1\right) \
\left(j_1-j_2+\bar{j}_1+\bar{j}_2+2\right) \
\left(j_1-j_2+\bar{j}_1+\bar{j}_2+3\right) \
\left(j_1-j_2+\bar{j}_1+\bar{j}_2+4\right) \
\left(-j_1+j_2+\bar{j}_1+\bar{j}_2+1\right) \
\left(j_1+j_2+\bar{j}_1+\bar{j}_2+3\right)}}{2 \sqrtpw{2} \
\sqrtpw{\left(j_1+1\right) \left(2 j_1+3\right) j_2 \left(2
j_2-1\right)}}
\eed,\bed%
\left(
\begin{array}{@{}c@{\;}c@{}||c@{\;}c@{\;\;}c@{\;}c@{}}
 \bar{j}_1 & \bar{j}_2+1 & \bar{j}_1 & \bar{j}_2 & 1 & 1 \\
 j_1+1 & j_2 & j_1 & j_2 & 1 & 1
\end{array}
\right)=\zfrac{\text{\Ncc}\left(\bar{j}_1,\bar{j}_2\right) \
\sqrtpw{\left(j_1-j_2-\bar{j}_1+\bar{j}_2\right) \
\left(j_1-j_2-\bar{j}_1+\bar{j}_2+1\right) \
\left(j_1+j_2-\bar{j}_1+\bar{j}_2+1\right) \
\left(j_1+j_2-\bar{j}_1+\bar{j}_2+2\right) \
\left(j_1-j_2+\bar{j}_1+\bar{j}_2+2\right) \
\left(j_1-j_2+\bar{j}_1+\bar{j}_2+3\right) \
\left(j_1+j_2+\bar{j}_1+\bar{j}_2+3\right) \
\left(j_1+j_2+\bar{j}_1+\bar{j}_2+4\right)} \left(-j_1^2+2
\bar{j}_2 \ j_1+j_2^2+\bar{j}_1^2-\bar{j}_2^2+j_2+2
\bar{j}_1+1\right)}{2 \ \sqrtpw{2} \sqrtpw{\left(2 j_1^2+5
j_1+3\right) j_2 \left(j_2+1\right)}}
\eed,\bed%
\left(
\begin{array}{@{}c@{\;}c@{}||c@{\;}c@{\;\;}c@{\;}c@{}}
 \bar{j}_1 & \bar{j}_2+1 & \bar{j}_1 & \bar{j}_2 & 1 & 1 \\
 j_1-1 & j_2 & j_1 & j_2 & 1 & 1
\end{array}
\right)=-\zfrac{\text{\Ncc}\left(\bar{j}_1,\bar{j}_2\right) \
\sqrtpw{\left(j_1-j_2+\bar{j}_1-\bar{j}_2-1\right) \
\left(j_1-j_2+\bar{j}_1-\bar{j}_2\right) \
\left(j_1+j_2+\bar{j}_1-\bar{j}_2\right) \
\left(j_1+j_2+\bar{j}_1-\bar{j}_2+1\right) \
\left(-j_1-j_2+\bar{j}_1+\bar{j}_2+1\right)
\left(-j_1-j_2+\bar{j}_1+\ \bar{j}_2+2\right)
\left(-j_1+j_2+\bar{j}_1+\bar{j}_2+2\right) \
\left(-j_1+j_2+\bar{j}_1+\bar{j}_2+3\right)} \left(j_1^2+2 \
\left(\bar{j}_2+1\right) j_1-j_2^2-\bar{j}_1^2+\bar{j}_2^2-j_2-2 \
\bar{j}_1+2 \bar{j}_2\right)}{2 \sqrtpw{2} \sqrtpw{j_1 \left(2
j_1-1\right) j_2 \left(j_2+1\right)}}
\eed,\bed%
\left(
\begin{array}{@{}c@{\;}c@{}||c@{\;}c@{\;\;}c@{\;}c@{}}
 \bar{j}_1 & \bar{j}_2+1 & \bar{j}_1 & \bar{j}_2 & 1 & 1 \\
 j_1 & j_2+1 & j_1 & j_2 & 1 & 1
\end{array}
\right)=\zfrac{\text{\Ncc}\left(\bar{j}_1,\bar{j}_2\right) \
\sqrtpw{\left(j_1-j_2+\bar{j}_1-\bar{j}_2-1\right) \
\left(j_1-j_2+\bar{j}_1-\bar{j}_2\right) \
\left(j_1+j_2-\bar{j}_1+\bar{j}_2+1\right) \
\left(j_1+j_2-\bar{j}_1+\bar{j}_2+2\right) \
\left(-j_1+j_2+\bar{j}_1+\bar{j}_2+2\right)
\left(-j_1+j_2+\bar{j}_1+\ \bar{j}_2+3\right)
\left(j_1+j_2+\bar{j}_1+\bar{j}_2+3\right) \
\left(j_1+j_2+\bar{j}_1+\bar{j}_2+4\right)} \
\left(j_1^2+j_1-j_2^2+\bar{j}_1^2-\bar{j}_2^2+2 \bar{j}_1+2 j_2 \
\bar{j}_2+1\right)}{2 \sqrtpw{2} \sqrtpw{j_1 \left(j_1+1\right)
\left(2 \ j_2^2+5 j_2+3\right)}}
\eed,\bed%
\left(
\begin{array}{@{}c@{\;}c@{}||c@{\;}c@{\;\;}c@{\;}c@{}}
 \bar{j}_1 & \bar{j}_2+1 & \bar{j}_1 & \bar{j}_2 & 1 & 1 \\
 j_1 & j_2-1 & j_1 & j_2 & 1 & 1
\end{array}
\right)=\zfrac{\text{\Ncc}\left(\bar{j}_1,\bar{j}_2\right) \
\sqrtpw{\left(-j_1+j_2+\bar{j}_1-\bar{j}_2-1\right) \
\left(-j_1+j_2+\bar{j}_1-\bar{j}_2\right) \
\left(j_1+j_2+\bar{j}_1-\bar{j}_2\right) \
\left(j_1+j_2+\bar{j}_1-\bar{j}_2+1\right) \
\left(-j_1-j_2+\bar{j}_1+\bar{j}_2+1\right)
\left(-j_1-j_2+\bar{j}_1+\ \bar{j}_2+2\right)
\left(j_1-j_2+\bar{j}_1+\bar{j}_2+2\right) \
\left(j_1-j_2+\bar{j}_1+\bar{j}_2+3\right)} \
\left(j_1^2+j_1-j_2^2+\bar{j}_1^2-\bar{j}_2^2+2 \bar{j}_1-2 \
\bar{j}_2-2 j_2 \left(\bar{j}_2+1\right)\right)}{2 \sqrtpw{2}
\sqrtpw{j_1 \ \left(j_1+1\right) j_2 \left(2 j_2-1\right)}}
\eed,\bed%
\left(
\begin{array}{@{}c@{\;}c@{}||c@{\;}c@{\;\;}c@{\;}c@{}}
 \bar{j}_1 & \bar{j}_2+1 & \bar{j}_1 & \bar{j}_2 & 1 & 1 \\
 j_1 & j_2 & j_1 & j_2 & 1 & 1
\end{array}
\right)=\zfrac{\text{\Ncc}\left(\bar{j}_1,\bar{j}_2\right) \
\sqrtpw{\left(j_1-j_2+\bar{j}_1-\bar{j}_2\right) \
\left(-j_1+j_2+\bar{j}_1-\bar{j}_2\right) \
\left(j_1+j_2+\bar{j}_1-\bar{j}_2+1\right) \
\left(j_1+j_2-\bar{j}_1+\bar{j}_2+1\right) \
\left(-j_1-j_2+\bar{j}_1+\bar{j}_2+1\right) \
\left(j_1-j_2+\bar{j}_1+\bar{j}_2+2\right) \
\left(-j_1+j_2+\bar{j}_1+\bar{j}_2+2\right) \
\left(j_1+j_2+\bar{j}_1+\bar{j}_2+3\right)} \
\left(j_1^2+j_1+j_2^2+\bar{j}_1^2-\bar{j}_2^2+j_2+2 \bar{j}_1-2 \
\bar{j}_2\right)}{2 \sqrtpw{2} \sqrtpw{j_1 \left(j_1+1\right) j_2
\ \left(j_2+1\right)}}
\eed,\bed%
\left(
\begin{array}{@{}c@{\;}c@{}||c@{\;}c@{\;\;}c@{\;}c@{}}
 \bar{j}_1 & \bar{j}_2+1 & \bar{j}_1 & \bar{j}_2 & 1 & 1 \\
 j_1+\frac{1}{2} & j_2+\frac{1}{2} & j_1 & j_2 & \frac{1}{2} & \
\frac{1}{2}
\end{array}
\right)=-\zfrac{\text{\Ncc}\left(\bar{j}_1,\bar{j}_2\right)
\left(2 \ j_1+2 j_2-2 \bar{j}_2+1\right) \
\sqrtpw{\left(j_1-j_2+\bar{j}_1-\bar{j}_2\right) \
\left(-j_1+j_2+\bar{j}_1-\bar{j}_2\right) \
\left(j_1+j_2-\bar{j}_1+\bar{j}_2+1\right) \
\left(j_1+j_2-\bar{j}_1+\bar{j}_2+2\right) \
\left(j_1-j_2+\bar{j}_1+\bar{j}_2+2\right) \
\left(-j_1+j_2+\bar{j}_1+\bar{j}_2+2\right) \
\left(j_1+j_2+\bar{j}_1+\bar{j}_2+3\right) \
\left(j_1+j_2+\bar{j}_1+\bar{j}_2+4\right)}}{2 \sqrtpw{2} \
\sqrtpw{\left(j_1+1\right) \left(j_2+1\right)}}
\eed,\bed%
\left(
\begin{array}{@{}c@{\;}c@{}||c@{\;}c@{\;\;}c@{\;}c@{}}
 \bar{j}_1 & \bar{j}_2+1 & \bar{j}_1 & \bar{j}_2 & 1 & 1 \\
 j_1-\frac{1}{2} & j_2-\frac{1}{2} & j_1 & j_2 & \frac{1}{2} & \
\frac{1}{2}
\end{array}
\right)=\zfrac{\text{\Ncc}\left(\bar{j}_1,\bar{j}_2\right) \
\sqrtpw{\left(j_1-j_2+\bar{j}_1-\bar{j}_2\right) \
\left(-j_1+j_2+\bar{j}_1-\bar{j}_2\right) \
\left(j_1+j_2+\bar{j}_1-\bar{j}_2\right) \
\left(j_1+j_2+\bar{j}_1-\bar{j}_2+1\right) \
\left(-j_1-j_2+\bar{j}_1+\bar{j}_2+1\right)
\left(-j_1-j_2+\bar{j}_1+\ \bar{j}_2+2\right)
\left(j_1-j_2+\bar{j}_1+\bar{j}_2+2\right) \
\left(-j_1+j_2+\bar{j}_1+\bar{j}_2+2\right)} \left(2 j_1+2 j_2+2 \
\bar{j}_2+3\right)}{2 \sqrtpw{2} \sqrtpw{j_1 j_2}}
\eed,\bed%
\left(
\begin{array}{@{}c@{\;}c@{}||c@{\;}c@{\;\;}c@{\;}c@{}}
 \bar{j}_1 & \bar{j}_2+1 & \bar{j}_1 & \bar{j}_2 & 1 & 1 \\
 j_1+\frac{1}{2} & j_2-\frac{1}{2} & j_1 & j_2 & \frac{1}{2} & \
\frac{1}{2}
\end{array}
\right)=\zfrac{\text{\Ncc}\left(\bar{j}_1,\bar{j}_2\right) \
\sqrtpw{\left(-j_1+j_2+\bar{j}_1-\bar{j}_2-1\right) \
\left(-j_1+j_2+\bar{j}_1-\bar{j}_2\right) \
\left(j_1+j_2+\bar{j}_1-\bar{j}_2+1\right) \
\left(j_1+j_2-\bar{j}_1+\bar{j}_2+1\right) \
\left(-j_1-j_2+\bar{j}_1+\bar{j}_2+1\right) \
\left(j_1-j_2+\bar{j}_1+\bar{j}_2+2\right) \
\left(j_1-j_2+\bar{j}_1+\bar{j}_2+3\right) \
\left(j_1+j_2+\bar{j}_1+\bar{j}_2+3\right)} \left(-2 j_1+2 j_2+2 \
\bar{j}_2+1\right)}{2 \sqrtpw{2} \sqrtpw{\left(j_1+1\right) j_2}}
\eed,\bed%
\left(
\begin{array}{@{}c@{\;}c@{}||c@{\;}c@{\;\;}c@{\;}c@{}}
 \bar{j}_1 & \bar{j}_2+1 & \bar{j}_1 & \bar{j}_2 & 1 & 1 \\
 j_1-\frac{1}{2} & j_2+\frac{1}{2} & j_1 & j_2 & \frac{1}{2} & \
\frac{1}{2}
\end{array}
\right)=\zfrac{\text{\Ncc}\left(\bar{j}_1,\bar{j}_2\right) \
\sqrtpw{\left(j_1-j_2+\bar{j}_1-\bar{j}_2-1\right) \
\left(j_1-j_2+\bar{j}_1-\bar{j}_2\right) \
\left(j_1+j_2+\bar{j}_1-\bar{j}_2+1\right) \
\left(j_1+j_2-\bar{j}_1+\bar{j}_2+1\right) \
\left(-j_1-j_2+\bar{j}_1+\bar{j}_2+1\right)
\left(-j_1+j_2+\bar{j}_1+\ \bar{j}_2+2\right)
\left(-j_1+j_2+\bar{j}_1+\bar{j}_2+3\right) \
\left(j_1+j_2+\bar{j}_1+\bar{j}_2+3\right)} \left(2 j_1-2 j_2+2 \
\bar{j}_2+1\right)}{2 \sqrtpw{2} \sqrtpw{j_1 \left(j_2+1\right)}}
\eed,\bed%
\left(
\begin{array}{@{}c@{\;}c@{}||c@{\;}c@{\;\;}c@{\;}c@{}}
 \bar{j}_1 & \bar{j}_2+1 & \bar{j}_1 & \bar{j}_2 & 1 & 1 \\
 j_1 & j_2 & j_1 & j_2 & 0 & 0
\end{array}
\right)=-\sqrtpw{\frac{5}{2}}
\text{\Ncc}\left(\bar{j}_1,\bar{j}_2\right) \
\sqrtpw{\left(j_1-j_2+\bar{j}_1-\bar{j}_2\right) \
\left(-j_1+j_2+\bar{j}_1-\bar{j}_2\right) \
\left(j_1+j_2+\bar{j}_1-\bar{j}_2+1\right) \
\left(j_1+j_2-\bar{j}_1+\bar{j}_2+1\right) \
\left(-j_1-j_2+\bar{j}_1+\bar{j}_2+1\right) \
\left(j_1-j_2+\bar{j}_1+\bar{j}_2+2\right) \
\left(-j_1+j_2+\bar{j}_1+\bar{j}_2+2\right) \
\left(j_1+j_2+\bar{j}_1+\bar{j}_2+3\right)} \eed
.

\vspace{1cm} The CG coefficients $(\bar{j}_1, \bar{j}_2) \otimes
(\bar 1, \bar 1) \rightarrow (\bar{j}_1+1, \bar{j}_2-1)$ are:

\bed \text{\Ndc}\left(\bar{j}_1,\bar{j}_2\right) = \left(2 \left(2
\bar{j}_1+2\right) \left(2 \bar{j}_1+3\right) \left(2 \bar{j}_1-2
\bar{j}_2+1\right) \left(2 \bar{j}_1-2
   \bar{j}_2+3\right) \left(\bar{j}_1-\bar{j}_2+1\right)
 \left(\bar{j}_1-\bar{j}_2+2\right) \bar{j}_2 \left(2
   \bar{j}_2+1\right)\right)^{-\frac 12} \eed

\bed \left(
\begin{array}{@{}c@{\;}c@{}||c@{\;}c@{\;\;}c@{\;}c@{}}
 \bar{j}_1+1 & \bar{j}_2-1 & \bar{j}_1 & \bar{j}_2 & 1 & 1 \\
 j_1+1 & j_2+1 & j_1 & j_2 & 1 & 1
\end{array}
\right)=\zfrac{\text{\Ndc}\left(\bar{j}_1,\bar{j}_2\right) \
\sqrtpw{\left(j_1-j_2+\bar{j}_1-\bar{j}_2+1\right) \
\left(j_1-j_2+\bar{j}_1-\bar{j}_2+2\right) \
\left(-j_1+j_2+\bar{j}_1-\bar{j}_2+1\right)
\left(-j_1+j_2+\bar{j}_1-\ \bar{j}_2+2\right)
\left(j_1+j_2+\bar{j}_1-\bar{j}_2+2\right) \
\left(j_1+j_2+\bar{j}_1-\bar{j}_2+3\right) \
\left(j_1+j_2+\bar{j}_1-\bar{j}_2+4\right) \
\left(j_1+j_2+\bar{j}_1-\bar{j}_2+5\right) \
\left(-j_1-j_2+\bar{j}_1+\bar{j}_2-1\right)
\left(-j_1-j_2+\bar{j}_1+\ \bar{j}_2\right)
\left(j_1+j_2+\bar{j}_1+\bar{j}_2+3\right) \
\left(j_1+j_2+\bar{j}_1+\bar{j}_2+4\right)}}{4
\sqrtpw{\left(j_1+1\right) \left(2 j_1+3\right) \left(j_2+1\right)
\left(2 j_2+3\right)}}
\eed,\bed%
\left(
\begin{array}{@{}c@{\;}c@{}||c@{\;}c@{\;\;}c@{\;}c@{}}
 \bar{j}_1+1 & \bar{j}_2-1 & \bar{j}_1 & \bar{j}_2 & 1 & 1 \\
 j_1-1 & j_2-1 & j_1 & j_2 & 1 & 1
\end{array}
\right)=\zfrac{\text{\Ndc}\left(\bar{j}_1,\bar{j}_2\right) \
\sqrtpw{\left(j_1-j_2+\bar{j}_1-\bar{j}_2+1\right) \
\left(j_1-j_2+\bar{j}_1-\bar{j}_2+2\right) \
\left(-j_1+j_2+\bar{j}_1-\bar{j}_2+1\right)
\left(-j_1+j_2+\bar{j}_1-\ \bar{j}_2+2\right)
\left(j_1+j_2-\bar{j}_1+\bar{j}_2-3\right) \
\left(j_1+j_2-\bar{j}_1+\bar{j}_2-2\right) \
\left(j_1+j_2-\bar{j}_1+\bar{j}_2-1\right) \
\left(j_1+j_2-\bar{j}_1+\bar{j}_2\right) \
\left(-j_1-j_2+\bar{j}_1+\bar{j}_2+1\right)
\left(-j_1-j_2+\bar{j}_1+\ \bar{j}_2+2\right)
\left(j_1+j_2+\bar{j}_1+\bar{j}_2+1\right) \
\left(j_1+j_2+\bar{j}_1+\bar{j}_2+2\right)}}{4 \sqrtpw{j_1 \left(2
\ j_1-1\right) j_2 \left(2 j_2-1\right)}}
\eed,\bed%
\left(
\begin{array}{@{}c@{\;}c@{}||c@{\;}c@{\;\;}c@{\;}c@{}}
 \bar{j}_1+1 & \bar{j}_2-1 & \bar{j}_1 & \bar{j}_2 & 1 & 1 \\
 j_1-1 & j_2+1 & j_1 & j_2 & 1 & 1
\end{array}
\right)=\zfrac{\text{\Ndc}\left(\bar{j}_1,\bar{j}_2\right) \
\sqrtpw{\left(-j_1+j_2+\bar{j}_1-\bar{j}_2+1\right) \
\left(-j_1+j_2+\bar{j}_1-\bar{j}_2+2\right)
\left(-j_1+j_2+\bar{j}_1-\ \bar{j}_2+3\right)
\left(-j_1+j_2+\bar{j}_1-\bar{j}_2+4\right) \
\left(j_1+j_2+\bar{j}_1-\bar{j}_2+2\right) \
\left(j_1+j_2+\bar{j}_1-\bar{j}_2+3\right) \
\left(j_1+j_2-\bar{j}_1+\bar{j}_2-1\right) \
\left(j_1+j_2-\bar{j}_1+\bar{j}_2\right) \
\left(j_1-j_2+\bar{j}_1+\bar{j}_2\right) \
\left(j_1-j_2+\bar{j}_1+\bar{j}_2+1\right) \
\left(-j_1+j_2+\bar{j}_1+\bar{j}_2+2\right)
\left(-j_1+j_2+\bar{j}_1+\ \bar{j}_2+3\right)}}{4 \sqrtpw{j_1
\left(2 j_1-1\right) \left(2 j_2^2+5 \ j_2+3\right)}}
\eed,\bed%
\left(
\begin{array}{@{}c@{\;}c@{}||c@{\;}c@{\;\;}c@{\;}c@{}}
 \bar{j}_1+1 & \bar{j}_2-1 & \bar{j}_1 & \bar{j}_2 & 1 & 1 \\
 j_1+1 & j_2-1 & j_1 & j_2 & 1 & 1
\end{array}
\right)=\zfrac{\text{\Ndc}\left(\bar{j}_1,\bar{j}_2\right) \
\sqrtpw{\left(j_1-j_2+\bar{j}_1-\bar{j}_2+1\right) \
\left(j_1-j_2+\bar{j}_1-\bar{j}_2+2\right) \
\left(j_1-j_2+\bar{j}_1-\bar{j}_2+3\right) \
\left(j_1-j_2+\bar{j}_1-\bar{j}_2+4\right) \
\left(j_1+j_2+\bar{j}_1-\bar{j}_2+2\right) \
\left(j_1+j_2+\bar{j}_1-\bar{j}_2+3\right) \
\left(j_1+j_2-\bar{j}_1+\bar{j}_2-1\right) \
\left(j_1+j_2-\bar{j}_1+\bar{j}_2\right) \
\left(j_1-j_2+\bar{j}_1+\bar{j}_2+2\right) \
\left(j_1-j_2+\bar{j}_1+\bar{j}_2+3\right) \
\left(-j_1+j_2+\bar{j}_1+\bar{j}_2\right) \
\left(-j_1+j_2+\bar{j}_1+\bar{j}_2+1\right)}}{4 \sqrtpw{\left(2
j_1^2+5 \ j_1+3\right) j_2 \left(2 j_2-1\right)}}
\eed,\bed%
\left(
\begin{array}{@{}c@{\;}c@{}||c@{\;}c@{\;\;}c@{\;}c@{}}
 \bar{j}_1+1 & \bar{j}_2-1 & \bar{j}_1 & \bar{j}_2 & 1 & 1 \\
 j_1+1 & j_2 & j_1 & j_2 & 1 & 1
\end{array}
\right)=-\zfrac{\text{\Ndc}\left(\bar{j}_1,\bar{j}_2\right) \
\sqrtpw{\left(j_1-j_2+\bar{j}_1-\bar{j}_2+1\right) \
\left(j_1-j_2+\bar{j}_1-\bar{j}_2+2\right) \
\left(j_1-j_2+\bar{j}_1-\bar{j}_2+3\right) \
\left(-j_1+j_2+\bar{j}_1-\bar{j}_2+1\right) \
\left(j_1+j_2+\bar{j}_1-\bar{j}_2+2\right) \
\left(j_1+j_2+\bar{j}_1-\bar{j}_2+3\right) \
\left(j_1+j_2+\bar{j}_1-\bar{j}_2+4\right) \
\left(j_1+j_2-\bar{j}_1+\bar{j}_2\right) \
\left(-j_1-j_2+\bar{j}_1+\bar{j}_2\right) \
\left(j_1-j_2+\bar{j}_1+\bar{j}_2+2\right) \
\left(-j_1+j_2+\bar{j}_1+\bar{j}_2+1\right) \
\left(j_1+j_2+\bar{j}_1+\bar{j}_2+3\right)}}{4 \sqrtpw{\left(2
j_1^2+5 \ j_1+3\right) j_2 \left(j_2+1\right)}}
\eed,\bed%
\left(
\begin{array}{@{}c@{\;}c@{}||c@{\;}c@{\;\;}c@{\;}c@{}}
 \bar{j}_1+1 & \bar{j}_2-1 & \bar{j}_1 & \bar{j}_2 & 1 & 1 \\
 j_1-1 & j_2 & j_1 & j_2 & 1 & 1
\end{array}
\right)=-\zfrac{\text{\Ndc}\left(\bar{j}_1,\bar{j}_2\right) \
\sqrtpw{\left(j_1-j_2+\bar{j}_1-\bar{j}_2+1\right) \
\left(-j_1+j_2+\bar{j}_1-\bar{j}_2+1\right)
\left(-j_1+j_2+\bar{j}_1-\ \bar{j}_2+2\right)
\left(-j_1+j_2+\bar{j}_1-\bar{j}_2+3\right) \
\left(j_1+j_2+\bar{j}_1-\bar{j}_2+2\right) \
\left(j_1+j_2-\bar{j}_1+\bar{j}_2-2\right) \
\left(j_1+j_2-\bar{j}_1+\bar{j}_2-1\right) \
\left(j_1+j_2-\bar{j}_1+\bar{j}_2\right) \
\left(-j_1-j_2+\bar{j}_1+\bar{j}_2+1\right) \
\left(j_1-j_2+\bar{j}_1+\bar{j}_2+1\right) \
\left(-j_1+j_2+\bar{j}_1+\bar{j}_2+2\right) \
\left(j_1+j_2+\bar{j}_1+\bar{j}_2+2\right)}}{4 \sqrtpw{j_1 \left(2
\ j_1-1\right) j_2 \left(j_2+1\right)}}
\eed,\bed%
\left(
\begin{array}{@{}c@{\;}c@{}||c@{\;}c@{\;\;}c@{\;}c@{}}
 \bar{j}_1+1 & \bar{j}_2-1 & \bar{j}_1 & \bar{j}_2 & 1 & 1 \\
 j_1 & j_2+1 & j_1 & j_2 & 1 & 1
\end{array}
\right)=-\zfrac{\text{\Ndc}\left(\bar{j}_1,\bar{j}_2\right) \
\sqrtpw{\left(j_1-j_2+\bar{j}_1-\bar{j}_2+1\right) \
\left(-j_1+j_2+\bar{j}_1-\bar{j}_2+1\right)
\left(-j_1+j_2+\bar{j}_1-\ \bar{j}_2+2\right)
\left(-j_1+j_2+\bar{j}_1-\bar{j}_2+3\right) \
\left(j_1+j_2+\bar{j}_1-\bar{j}_2+2\right) \
\left(j_1+j_2+\bar{j}_1-\bar{j}_2+3\right) \
\left(j_1+j_2+\bar{j}_1-\bar{j}_2+4\right) \
\left(j_1+j_2-\bar{j}_1+\bar{j}_2\right) \
\left(-j_1-j_2+\bar{j}_1+\bar{j}_2\right) \
\left(j_1-j_2+\bar{j}_1+\bar{j}_2+1\right) \
\left(-j_1+j_2+\bar{j}_1+\bar{j}_2+2\right) \
\left(j_1+j_2+\bar{j}_1+\bar{j}_2+3\right)}}{4 \sqrtpw{j_1
\left(j_1+1\right) \left(2 j_2^2+5 j_2+3\right)}}
\eed,\bed%
\left(
\begin{array}{@{}c@{\;}c@{}||c@{\;}c@{\;\;}c@{\;}c@{}}
 \bar{j}_1+1 & \bar{j}_2-1 & \bar{j}_1 & \bar{j}_2 & 1 & 1 \\
 j_1 & j_2-1 & j_1 & j_2 & 1 & 1
\end{array}
\right)=-\zfrac{\text{\Ndc}\left(\bar{j}_1,\bar{j}_2\right) \
\sqrtpw{\left(j_1-j_2+\bar{j}_1-\bar{j}_2+1\right) \
\left(j_1-j_2+\bar{j}_1-\bar{j}_2+2\right) \
\left(j_1-j_2+\bar{j}_1-\bar{j}_2+3\right) \
\left(-j_1+j_2+\bar{j}_1-\bar{j}_2+1\right) \
\left(j_1+j_2+\bar{j}_1-\bar{j}_2+2\right) \
\left(j_1+j_2-\bar{j}_1+\bar{j}_2-2\right) \
\left(j_1+j_2-\bar{j}_1+\bar{j}_2-1\right) \
\left(j_1+j_2-\bar{j}_1+\bar{j}_2\right) \
\left(-j_1-j_2+\bar{j}_1+\bar{j}_2+1\right) \
\left(j_1-j_2+\bar{j}_1+\bar{j}_2+2\right) \
\left(-j_1+j_2+\bar{j}_1+\bar{j}_2+1\right) \
\left(j_1+j_2+\bar{j}_1+\bar{j}_2+2\right)}}{4 \sqrtpw{j_1
\left(j_1+1\right) j_2 \left(2 j_2-1\right)}}
\eed,\bed%
\left(
\begin{array}{@{}c@{\;}c@{}||c@{\;}c@{\;\;}c@{\;}c@{}}
 \bar{j}_1+1 & \bar{j}_2-1 & \bar{j}_1 & \bar{j}_2 & 1 & 1 \\
 j_1 & j_2 & j_1 & j_2 & 1 & 1
\end{array}
\right)=\zfrac{\text{\Ndc}\left(\bar{j}_1,\bar{j}_2\right) \
\sqrtpw{\left(j_1-j_2+\bar{j}_1-\bar{j}_2+1\right) \
\left(j_1-j_2+\bar{j}_1-\bar{j}_2+2\right) \
\left(-j_1+j_2+\bar{j}_1-\bar{j}_2+1\right)
\left(-j_1+j_2+\bar{j}_1-\ \bar{j}_2+2\right)
\left(j_1+j_2+\bar{j}_1-\bar{j}_2+2\right) \
\left(j_1+j_2+\bar{j}_1-\bar{j}_2+3\right) \
\left(j_1+j_2-\bar{j}_1+\bar{j}_2-1\right) \
\left(j_1+j_2-\bar{j}_1+\bar{j}_2\right)} \left(2 j_1 \
j_2+\left(-j_1-j_2+\bar{j}_1+\bar{j}_2+1\right) \
\left(j_1+j_2+\bar{j}_1+\bar{j}_2+2\right)\right)}{4 \sqrtpw{j_1 \
\left(j_1+1\right) j_2 \left(j_2+1\right)}}
\eed,\bed%
\left(
\begin{array}{@{}c@{\;}c@{}||c@{\;}c@{\;\;}c@{\;}c@{}}
 \bar{j}_1+1 & \bar{j}_2-1 & \bar{j}_1 & \bar{j}_2 & 1 & 1 \\
 j_1+\frac{1}{2} & j_2+\frac{1}{2} & j_1 & j_2 & \frac{1}{2} & \
\frac{1}{2}
\end{array}
\right)=-\zfrac{\text{\Ndc}\left(\bar{j}_1,\bar{j}_2\right) \
\sqrtpw{\left(j_1-j_2+\bar{j}_1-\bar{j}_2+1\right) \
\left(j_1-j_2+\bar{j}_1-\bar{j}_2+2\right) \
\left(-j_1+j_2+\bar{j}_1-\bar{j}_2+1\right)
\left(-j_1+j_2+\bar{j}_1-\ \bar{j}_2+2\right)
\left(j_1+j_2+\bar{j}_1-\bar{j}_2+2\right) \
\left(j_1+j_2+\bar{j}_1-\bar{j}_2+3\right) \
\left(j_1+j_2+\bar{j}_1-\bar{j}_2+4\right) \
\left(j_1+j_2-\bar{j}_1+\bar{j}_2\right) \
\left(-j_1-j_2+\bar{j}_1+\bar{j}_2\right) \
\left(j_1+j_2+\bar{j}_1+\bar{j}_2+3\right)}}{2
\sqrtpw{\left(j_1+1\right) \left(j_2+1\right)}}
\eed,\bed%
\left(
\begin{array}{@{}c@{\;}c@{}||c@{\;}c@{\;\;}c@{\;}c@{}}
 \bar{j}_1+1 & \bar{j}_2-1 & \bar{j}_1 & \bar{j}_2 & 1 & 1 \\
 j_1-\frac{1}{2} & j_2-\frac{1}{2} & j_1 & j_2 & \frac{1}{2} & \
\frac{1}{2}
\end{array}
\right)=-\zfrac{\text{\Ndc}\left(\bar{j}_1,\bar{j}_2\right) \
\sqrtpw{\left(j_1-j_2+\bar{j}_1-\bar{j}_2+1\right) \
\left(j_1-j_2+\bar{j}_1-\bar{j}_2+2\right) \
\left(-j_1+j_2+\bar{j}_1-\bar{j}_2+1\right)
\left(-j_1+j_2+\bar{j}_1-\ \bar{j}_2+2\right)
\left(j_1+j_2+\bar{j}_1-\bar{j}_2+2\right) \
\left(j_1+j_2-\bar{j}_1+\bar{j}_2-2\right) \
\left(j_1+j_2-\bar{j}_1+\bar{j}_2-1\right) \
\left(j_1+j_2-\bar{j}_1+\bar{j}_2\right) \
\left(-j_1-j_2+\bar{j}_1+\bar{j}_2+1\right) \
\left(j_1+j_2+\bar{j}_1+\bar{j}_2+2\right)}}{2 \sqrtpw{j_1 j_2}}
\eed,\bed%
\left(
\begin{array}{@{}c@{\;}c@{}||c@{\;}c@{\;\;}c@{\;}c@{}}
 \bar{j}_1+1 & \bar{j}_2-1 & \bar{j}_1 & \bar{j}_2 & 1 & 1 \\
 j_1+\frac{1}{2} & j_2-\frac{1}{2} & j_1 & j_2 & \frac{1}{2} & \
\frac{1}{2}
\end{array}
\right)=\zfrac{\text{\Ndc}\left(\bar{j}_1,\bar{j}_2\right) \
\sqrtpw{\left(j_1-j_2+\bar{j}_1-\bar{j}_2+1\right) \
\left(j_1-j_2+\bar{j}_1-\bar{j}_2+2\right) \
\left(j_1-j_2+\bar{j}_1-\bar{j}_2+3\right) \
\left(-j_1+j_2+\bar{j}_1-\bar{j}_2+1\right) \
\left(j_1+j_2+\bar{j}_1-\bar{j}_2+2\right) \
\left(j_1+j_2+\bar{j}_1-\bar{j}_2+3\right) \
\left(j_1+j_2-\bar{j}_1+\bar{j}_2-1\right) \
\left(j_1+j_2-\bar{j}_1+\bar{j}_2\right) \
\left(j_1-j_2+\bar{j}_1+\bar{j}_2+2\right) \
\left(-j_1+j_2+\bar{j}_1+\bar{j}_2+1\right)}}{2
\sqrtpw{\left(j_1+1\right) j_2}}
\eed,\bed%
\left(
\begin{array}{@{}c@{\;}c@{}||c@{\;}c@{\;\;}c@{\;}c@{}}
 \bar{j}_1+1 & \bar{j}_2-1 & \bar{j}_1 & \bar{j}_2 & 1 & 1 \\
 j_1-\frac{1}{2} & j_2+\frac{1}{2} & j_1 & j_2 & \frac{1}{2} & \
\frac{1}{2}
\end{array}
\right)=\zfrac{\text{\Ndc}\left(\bar{j}_1,\bar{j}_2\right) \
\sqrtpw{\left(j_1-j_2+\bar{j}_1-\bar{j}_2+1\right) \
\left(-j_1+j_2+\bar{j}_1-\bar{j}_2+1\right)
\left(-j_1+j_2+\bar{j}_1-\ \bar{j}_2+2\right)
\left(-j_1+j_2+\bar{j}_1-\bar{j}_2+3\right) \
\left(j_1+j_2+\bar{j}_1-\bar{j}_2+2\right) \
\left(j_1+j_2+\bar{j}_1-\bar{j}_2+3\right) \
\left(j_1+j_2-\bar{j}_1+\bar{j}_2-1\right) \
\left(j_1+j_2-\bar{j}_1+\bar{j}_2\right) \
\left(j_1-j_2+\bar{j}_1+\bar{j}_2+1\right) \
\left(-j_1+j_2+\bar{j}_1+\bar{j}_2+2\right)}}{2 \sqrtpw{j_1
\left(j_2+1\ \right)}}
\eed,\bed%
\left(
\begin{array}{@{}c@{\;}c@{}||c@{\;}c@{\;\;}c@{\;}c@{}}
 \bar{j}_1+1 & \bar{j}_2-1 & \bar{j}_1 & \bar{j}_2 & 1 & 1 \\
 j_1 & j_2 & j_1 & j_2 & 0 & 0
\end{array}
\right)=\frac{1}{2} \sqrtpw{5}
\text{\Ndc}\left(\bar{j}_1,\bar{j}_2\right)
\sqrtpw{\left(j_1-j_2+\bar{j}_1-\bar{j}_2+1\right) \left(j_1-j_2+\
\bar{j}_1-\bar{j}_2+2\right)
\left(-j_1+j_2+\bar{j}_1-\bar{j}_2+1\right)
\left(-j_1+j_2+\bar{j}_1-\bar{j}_2+2\right) \
\left(j_1+j_2+\bar{j}_1-\bar{j}_2+2\right) \
\left(j_1+j_2+\bar{j}_1-\bar{j}_2+3\right) \
\left(j_1+j_2-\bar{j}_1+\bar{j}_2-1\right) \
\left(j_1+j_2-\bar{j}_1+\bar{j}_2\right)}
\eed%
.

\vspace{1cm} The CG coefficients $(\bar{j}_1, \bar{j}_2) \otimes
(\bar 1, \bar 1) \rightarrow (\bar{j}_1+\frac 12, \bar{j}_2
+\frac12)$ are:

\bed \text{\Nec}\left(\bar{j}_1,\bar{j}_2\right) = \left(\left(2
\bar{j}_1+2\right) \left(\bar{j}_1-\bar{j}_2\right)
\left(\bar{j}_1-\bar{j}_2+1\right)
\left(\bar{j}_1+\bar{j}_2+1\right)
   \left(\bar{j}_1+\bar{j}_2+2\right) \left(\bar{j}_1+\bar{j}_2+3\right)
   \left(2 \bar{j}_2+1\right) \left(2 \bar{j}_1+2
     \bar{j}_2+3\right)\right)^{-\frac 12} \eed

\bed \left(
\begin{array}{@{}c@{\;}c@{}||c@{\;}c@{\;\;}c@{\;}c@{}}
 \bar{j}_1+\frac{1}{2} & \bar{j}_2+\frac{1}{2} & \bar{j}_1 & \
\bar{j}_2 & 1 & 1 \\
 j_1+1 & j_2+1 & j_1 & j_2 & 1 & 1
\end{array}
\right)=-\zfrac{\text{\Nec}\left(\bar{j}_1,\bar{j}_2\right) \
\left(j_1-j_2\right)
\sqrtpw{\left(j_1+j_2+\bar{j}_1-\bar{j}_2+2\right) \
\left(j_1+j_2+\bar{j}_1-\bar{j}_2+3\right) \
\left(j_1+j_2-\bar{j}_1+\bar{j}_2+1\right) \
\left(j_1+j_2-\bar{j}_1+\bar{j}_2+2\right) \
\left(-j_1-j_2+\bar{j}_1+\bar{j}_2\right) \
\left(j_1-j_2+\bar{j}_1+\bar{j}_2+2\right) \
\left(-j_1+j_2+\bar{j}_1+\bar{j}_2+2\right) \
\left(j_1+j_2+\bar{j}_1+\bar{j}_2+3\right) \
\left(j_1+j_2+\bar{j}_1+\bar{j}_2+4\right) \
\left(j_1+j_2+\bar{j}_1+\bar{j}_2+5\right)}}{2 \sqrtpw{2} \
\sqrtpw{\left(j_1+1\right) \left(2 j_1+3\right) \left(j_2+1\right)
\ \left(2 j_2+3\right)}}
\eed,\bed%
\left(
\begin{array}{@{}c@{\;}c@{}||c@{\;}c@{\;\;}c@{\;}c@{}}
 \bar{j}_1+\frac{1}{2} & \bar{j}_2+\frac{1}{2} & \bar{j}_1 & \
\bar{j}_2 & 1 & 1 \\
 j_1-1 & j_2-1 & j_1 & j_2 & 1 & 1
\end{array}
\right)=\zfrac{\text{\Nec}\left(\bar{j}_1,\bar{j}_2\right)
\left(j_1-j_2\ \right)
\sqrtpw{\left(j_1+j_2+\bar{j}_1-\bar{j}_2\right) \left(j_1+j_2+\
\bar{j}_1-\bar{j}_2+1\right)
\left(j_1+j_2-\bar{j}_1+\bar{j}_2-1\right)
\left(j_1+j_2-\bar{j}_1+\bar{j}_2\right) \
\left(-j_1-j_2+\bar{j}_1+\bar{j}_2+1\right)
\left(-j_1-j_2+\bar{j}_1+\ \bar{j}_2+2\right)
\left(-j_1-j_2+\bar{j}_1+\bar{j}_2+3\right) \
\left(j_1-j_2+\bar{j}_1+\bar{j}_2+2\right) \
\left(-j_1+j_2+\bar{j}_1+\bar{j}_2+2\right) \
\left(j_1+j_2+\bar{j}_1+\bar{j}_2+2\right)}}{2 \sqrtpw{2}
\sqrtpw{j_1 \ \left(2 j_1-1\right) j_2 \left(2 j_2-1\right)}}
\eed,\bed%
\left(
\begin{array}{@{}c@{\;}c@{}||c@{\;}c@{\;\;}c@{\;}c@{}}
 \bar{j}_1+\frac{1}{2} & \bar{j}_2+\frac{1}{2} & \bar{j}_1 & \
\bar{j}_2 & 1 & 1 \\
 j_1-1 & j_2+1 & j_1 & j_2 & 1 & 1
\end{array}
\right)=\zfrac{\text{\Nec}\left(\bar{j}_1,\bar{j}_2\right) \
\left(j_1+j_2+1\right)
\sqrtpw{\left(j_1-j_2+\bar{j}_1-\bar{j}_2-1\right)
\left(j_1-j_2+\bar{j}_1-\bar{j}_2\right) \
\left(-j_1+j_2+\bar{j}_1-\bar{j}_2+1\right)
\left(-j_1+j_2+\bar{j}_1-\ \bar{j}_2+2\right)
\left(-j_1-j_2+\bar{j}_1+\bar{j}_2+1\right) \
\left(j_1-j_2+\bar{j}_1+\bar{j}_2+1\right) \
\left(-j_1+j_2+\bar{j}_1+\bar{j}_2+2\right)
\left(-j_1+j_2+\bar{j}_1+\ \bar{j}_2+3\right)
\left(-j_1+j_2+\bar{j}_1+\bar{j}_2+4\right) \
\left(j_1+j_2+\bar{j}_1+\bar{j}_2+3\right)}}{2 \sqrtpw{2}
\sqrtpw{j_1 \ \left(2 j_1-1\right) \left(2 j_2^2+5 j_2+3\right)}}
\eed,\bed%
\left(
\begin{array}{@{}c@{\;}c@{}||c@{\;}c@{\;\;}c@{\;}c@{}}
 \bar{j}_1+\frac{1}{2} & \bar{j}_2+\frac{1}{2} & \bar{j}_1 & \
\bar{j}_2 & 1 & 1 \\
 j_1+1 & j_2-1 & j_1 & j_2 & 1 & 1
\end{array}
\right)=-\zfrac{\text{\Nec}\left(\bar{j}_1,\bar{j}_2\right) \
\left(j_1+j_2+1\right)
\sqrtpw{\left(j_1-j_2+\bar{j}_1-\bar{j}_2+1\right)
\left(j_1-j_2+\bar{j}_1-\bar{j}_2+2\right) \
\left(-j_1+j_2+\bar{j}_1-\bar{j}_2-1\right)
\left(-j_1+j_2+\bar{j}_1-\ \bar{j}_2\right)
\left(-j_1-j_2+\bar{j}_1+\bar{j}_2+1\right) \
\left(j_1-j_2+\bar{j}_1+\bar{j}_2+2\right) \
\left(j_1-j_2+\bar{j}_1+\bar{j}_2+3\right) \
\left(j_1-j_2+\bar{j}_1+\bar{j}_2+4\right) \
\left(-j_1+j_2+\bar{j}_1+\bar{j}_2+1\right) \
\left(j_1+j_2+\bar{j}_1+\bar{j}_2+3\right)}}{2 \sqrtpw{2}
\sqrtpw{\left(2 \ j_1^2+5 j_1+3\right) j_2 \left(2 j_2-1\right)}}
\eed,\bed%
\left(
\begin{array}{@{}c@{\;}c@{}||c@{\;}c@{\;\;}c@{\;}c@{}}
 \bar{j}_1+\frac{1}{2} & \bar{j}_2+\frac{1}{2} & \bar{j}_1 & \
\bar{j}_2 & 1 & 1 \\
 j_1+1 & j_2 & j_1 & j_2 & 1 & 1
\end{array}
\right)=-\zfrac{\text{\Nec}\left(\bar{j}_1,\bar{j}_2\right) \
\sqrtpw{\left(j_1-j_2+\bar{j}_1-\bar{j}_2+1\right) \
\left(-j_1+j_2+\bar{j}_1-\bar{j}_2\right) \
\left(j_1+j_2+\bar{j}_1-\bar{j}_2+2\right) \
\left(j_1+j_2-\bar{j}_1+\bar{j}_2+1\right) \
\left(j_1-j_2+\bar{j}_1+\bar{j}_2+2\right) \
\left(j_1-j_2+\bar{j}_1+\bar{j}_2+3\right) \
\left(j_1+j_2+\bar{j}_1+\bar{j}_2+3\right) \
\left(j_1+j_2+\bar{j}_1+\bar{j}_2+4\right)} \left(j_2
\left(j_2+1\right)+\left(j_1+1\right)
\left(-j_1+\bar{j}_1+\bar{j}_2+1\right)\right)}{2 \sqrtpw{2}
\sqrtpw{\left(j_1+1\right) \left(2 j_1+3\right) j_2 \
\left(j_2+1\right)}}
\eed,\bed%
\left(
\begin{array}{@{}c@{\;}c@{}||c@{\;}c@{\;\;}c@{\;}c@{}}
 \bar{j}_1+\frac{1}{2} & \bar{j}_2+\frac{1}{2} & \bar{j}_1 & \
\bar{j}_2 & 1 & 1 \\
 j_1-1 & j_2 & j_1 & j_2 & 1 & 1
\end{array}
\right)=-\zfrac{\text{\Nec}\left(\bar{j}_1,\bar{j}_2\right) \
\sqrtpw{\left(j_1-j_2+\bar{j}_1-\bar{j}_2\right) \
\left(-j_1+j_2+\bar{j}_1-\bar{j}_2+1\right) \
\left(j_1+j_2+\bar{j}_1-\bar{j}_2+1\right) \
\left(j_1+j_2-\bar{j}_1+\bar{j}_2\right) \
\left(-j_1-j_2+\bar{j}_1+\bar{j}_2+1\right)
\left(-j_1-j_2+\bar{j}_1+\ \bar{j}_2+2\right)
\left(-j_1+j_2+\bar{j}_1+\bar{j}_2+2\right) \
\left(-j_1+j_2+\bar{j}_1+\bar{j}_2+3\right)} \
\left(j_1^2+\left(\bar{j}_1+\bar{j}_2+2\right) j_1-j_2
\left(j_2+1\right)\right)}{2 \sqrtpw{2} \sqrtpw{j_1 \left(2
j_1-1\right) j_2 \ \left(j_2+1\right)}}
\eed,\bed%
\left(
\begin{array}{@{}c@{\;}c@{}||c@{\;}c@{\;\;}c@{\;}c@{}}
 \bar{j}_1+\frac{1}{2} & \bar{j}_2+\frac{1}{2} & \bar{j}_1 & \
\bar{j}_2 & 1 & 1 \\
 j_1 & j_2+1 & j_1 & j_2 & 1 & 1
\end{array}
\right)=\zfrac{\text{\Nec}\left(\bar{j}_1,\bar{j}_2\right) \
\sqrtpw{\left(j_1-j_2+\bar{j}_1-\bar{j}_2\right) \
\left(-j_1+j_2+\bar{j}_1-\bar{j}_2+1\right) \
\left(j_1+j_2+\bar{j}_1-\bar{j}_2+2\right) \
\left(j_1+j_2-\bar{j}_1+\bar{j}_2+1\right) \
\left(-j_1+j_2+\bar{j}_1+\bar{j}_2+2\right)
\left(-j_1+j_2+\bar{j}_1+\ \bar{j}_2+3\right)
\left(j_1+j_2+\bar{j}_1+\bar{j}_2+3\right) \
\left(j_1+j_2+\bar{j}_1+\bar{j}_2+4\right)} \left(j_1
\left(j_1+1\right)+\left(j_2+1\right)
\left(-j_2+\bar{j}_1+\bar{j}_2+1\right)\right)}{2 \sqrtpw{2}
\sqrtpw{j_1 \left(j_1+1\right) \left(j_2+1\right) \ \left(2
j_2+3\right)}}
\eed,\bed%
\left(
\begin{array}{@{}c@{\;}c@{}||c@{\;}c@{\;\;}c@{\;}c@{}}
 \bar{j}_1+\frac{1}{2} & \bar{j}_2+\frac{1}{2} & \bar{j}_1 & \
\bar{j}_2 & 1 & 1 \\
 j_1 & j_2-1 & j_1 & j_2 & 1 & 1
\end{array}
\right)=-\zfrac{\text{\Nec}\left(\bar{j}_1,\bar{j}_2\right) \
\sqrtpw{\left(j_1-j_2+\bar{j}_1-\bar{j}_2+1\right) \
\left(-j_1+j_2+\bar{j}_1-\bar{j}_2\right) \
\left(j_1+j_2+\bar{j}_1-\bar{j}_2+1\right) \
\left(j_1+j_2-\bar{j}_1+\bar{j}_2\right) \
\left(-j_1-j_2+\bar{j}_1+\bar{j}_2+1\right)
\left(-j_1-j_2+\bar{j}_1+\ \bar{j}_2+2\right)
\left(j_1-j_2+\bar{j}_1+\bar{j}_2+2\right) \
\left(j_1-j_2+\bar{j}_1+\bar{j}_2+3\right)} \left(j_1^2+j_1-j_2 \
\left(j_2+\bar{j}_1+\bar{j}_2+2\right)\right)}{2 \sqrtpw{2}
\sqrtpw{j_1 \ \left(j_1+1\right) j_2 \left(2 j_2-1\right)}}
\eed,\bed%
\left(
\begin{array}{@{}c@{\;}c@{}||c@{\;}c@{\;\;}c@{\;}c@{}}
 \bar{j}_1+\frac{1}{2} & \bar{j}_2+\frac{1}{2} & \bar{j}_1 & \
\bar{j}_2 & 1 & 1 \\
 j_1 & j_2 & j_1 & j_2 & 1 & 1
\end{array}
\right)=\zfrac{\text{\Nec}\left(\bar{j}_1,\bar{j}_2\right)
\left(j_1-j_2\ \right) \left(j_1+j_2+1\right) \
\left(-j_1^2-j_1-j_2^2-j_2+\left(\bar{j}_1-\bar{j}_2\right) \
\left(\bar{j}_1-\bar{j}_2+1\right)\right) \
\sqrtpw{\left(-j_1-j_2+\bar{j}_1+\bar{j}_2+1\right) \
\left(j_1-j_2+\bar{j}_1+\bar{j}_2+2\right) \
\left(-j_1+j_2+\bar{j}_1+\bar{j}_2+2\right) \
\left(j_1+j_2+\bar{j}_1+\bar{j}_2+3\right)}}{2 \sqrtpw{2}
\sqrtpw{j_1 \ \left(j_1+1\right) j_2 \left(j_2+1\right)}}
\eed,\bed%
\left(
\begin{array}{@{}c@{\;}c@{}||c@{\;}c@{\;\;}c@{\;}c@{}}
 \bar{j}_1+\frac{1}{2} & \bar{j}_2+\frac{1}{2} & \bar{j}_1 & \
\bar{j}_2 & 1 & 1 \\
 j_1+\frac{1}{2} & j_2+\frac{1}{2} & j_1 & j_2 & \frac{1}{2} & \
\frac{1}{2}
\end{array}
\right)=\zfrac{\text{\Nec}\left(\bar{j}_1,\bar{j}_2\right)
\left(j_1-j_2\ \right) \left(2 j_1+2
j_2-\bar{j}_1-\bar{j}_2+1\right) \
\sqrtpw{\left(j_1+j_2+\bar{j}_1-\bar{j}_2+2\right) \
\left(j_1+j_2-\bar{j}_1+\bar{j}_2+1\right) \
\left(j_1-j_2+\bar{j}_1+\bar{j}_2+2\right) \
\left(-j_1+j_2+\bar{j}_1+\bar{j}_2+2\right) \
\left(j_1+j_2+\bar{j}_1+\bar{j}_2+3\right) \
\left(j_1+j_2+\bar{j}_1+\bar{j}_2+4\right)}}{2 \sqrtpw{2} \
\sqrtpw{\left(j_1+1\right) \left(j_2+1\right)}}
\eed,\bed%
\left(
\begin{array}{@{}c@{\;}c@{}||c@{\;}c@{\;\;}c@{\;}c@{}}
 \bar{j}_1+\frac{1}{2} & \bar{j}_2+\frac{1}{2} & \bar{j}_1 & \
\bar{j}_2 & 1 & 1 \\
 j_1-\frac{1}{2} & j_2-\frac{1}{2} & j_1 & j_2 & \frac{1}{2} & \
\frac{1}{2}
\end{array}
\right)=-\zfrac{\text{\Nec}\left(\bar{j}_1,\bar{j}_2\right) \
\left(j_1-j_2\right)
\sqrtpw{\left(j_1+j_2+\bar{j}_1-\bar{j}_2+1\right) \
\left(j_1+j_2-\bar{j}_1+\bar{j}_2\right) \
\left(-j_1-j_2+\bar{j}_1+\bar{j}_2+1\right)
\left(-j_1-j_2+\bar{j}_1+\ \bar{j}_2+2\right)
\left(j_1-j_2+\bar{j}_1+\bar{j}_2+2\right) \
\left(-j_1+j_2+\bar{j}_1+\bar{j}_2+2\right)} \left(2 j_1+2 \
j_2+\bar{j}_1+\bar{j}_2+3\right)}{2 \sqrtpw{2} \sqrtpw{j_1 j_2}}
\eed,\bed%
\left(
\begin{array}{@{}c@{\;}c@{}||c@{\;}c@{\;\;}c@{\;}c@{}}
 \bar{j}_1+\frac{1}{2} & \bar{j}_2+\frac{1}{2} & \bar{j}_1 & \
\bar{j}_2 & 1 & 1 \\
 j_1+\frac{1}{2} & j_2-\frac{1}{2} & j_1 & j_2 & \frac{1}{2} & \
\frac{1}{2}
\end{array}
\right)=-\zfrac{\text{\Nec}\left(\bar{j}_1,\bar{j}_2\right) \
\left(j_1+j_2+1\right)
\sqrtpw{\left(j_1-j_2+\bar{j}_1-\bar{j}_2+1\right)
\left(-j_1+j_2+\bar{j}_1-\bar{j}_2\right) \
\left(-j_1-j_2+\bar{j}_1+\bar{j}_2+1\right) \
\left(j_1-j_2+\bar{j}_1+\bar{j}_2+2\right) \
\left(j_1-j_2+\bar{j}_1+\bar{j}_2+3\right) \
\left(j_1+j_2+\bar{j}_1+\bar{j}_2+3\right)} \left(-2 j_1+2 \
j_2+\bar{j}_1+\bar{j}_2+1\right)}{2 \sqrtpw{2}
\sqrtpw{\left(j_1+1\right) \ j_2}}
\eed,\bed%
\left(
\begin{array}{@{}c@{\;}c@{}||c@{\;}c@{\;\;}c@{\;}c@{}}
 \bar{j}_1+\frac{1}{2} & \bar{j}_2+\frac{1}{2} & \bar{j}_1 & \
\bar{j}_2 & 1 & 1 \\
 j_1-\frac{1}{2} & j_2+\frac{1}{2} & j_1 & j_2 & \frac{1}{2} & \
\frac{1}{2}
\end{array}
\right)=\zfrac{\text{\Nec}\left(\bar{j}_1,\bar{j}_2\right) \
\left(j_1+j_2+1\right) \left(2 j_1-2
j_2+\bar{j}_1+\bar{j}_2+1\right) \
\sqrtpw{\left(j_1-j_2+\bar{j}_1-\bar{j}_2\right) \
\left(-j_1+j_2+\bar{j}_1-\bar{j}_2+1\right)
\left(-j_1-j_2+\bar{j}_1+\ \bar{j}_2+1\right)
\left(-j_1+j_2+\bar{j}_1+\bar{j}_2+2\right) \
\left(-j_1+j_2+\bar{j}_1+\bar{j}_2+3\right) \
\left(j_1+j_2+\bar{j}_1+\bar{j}_2+3\right)}}{2 \sqrtpw{2}
\sqrtpw{j_1 \ \left(j_2+1\right)}}
\eed,\bed%
\left(
\begin{array}{@{}c@{\;}c@{}||c@{\;}c@{\;\;}c@{\;}c@{}}
 \bar{j}_1+\frac{1}{2} & \bar{j}_2+\frac{1}{2} & \bar{j}_1 & \
\bar{j}_2 & 1 & 1 \\
 j_1 & j_2 & j_1 & j_2 & 0 & 0
\end{array}
\right)=\sqrtpw{\frac{5}{2}}
\text{\Nec}\left(\bar{j}_1,\bar{j}_2\right) \ \left(j_1-j_2\right)
\left(j_1+j_2+1\right) \
\sqrtpw{\left(-j_1-j_2+\bar{j}_1+\bar{j}_2+1\right) \
\left(j_1-j_2+\bar{j}_1+\bar{j}_2+2\right) \
\left(-j_1+j_2+\bar{j}_1+\bar{j}_2+2\right) \
\left(j_1+j_2+\bar{j}_1+\bar{j}_2+3\right)} \eed%
.

\vspace{1cm} The CG coefficients $(\bar{j}_1, \bar{j}_2) \otimes
(\bar 1, \bar 1) \rightarrow (\bar{j}_1+\frac 12, \bar{j}_2
-\frac12)$ are:

\bed \text{\Nfc}\left(\bar{j}_1,\bar{j}_2\right) = \left(\left(2
\bar{j}_1+2\right) \left(2 \bar{j}_1-2 \bar{j}_2+1\right)
\left(\bar{j}_1-\bar{j}_2\right)
   \left(\bar{j}_1-\bar{j}_2+1\right) \left(\bar{j}_1-\bar{j}_2+2\right)
   \left(\bar{j}_1+\bar{j}_2+1\right) \left(\bar{j}_1+\bar{j}_2+2\right)
   \left(2 \bar{j}_2+1\right)\right)^{-\frac 12} \eed

\bed \left(
\begin{array}{@{}c@{\;}c@{}||c@{\;}c@{\;\;}c@{\;}c@{}}
 \bar{j}_1+\frac{1}{2} & \bar{j}_2-\frac{1}{2} & \bar{j}_1 & \
\bar{j}_2 & 1 & 1 \\
 j_1+1 & j_2+1 & j_1 & j_2 & 1 & 1
\end{array}
\right)=\zfrac{\text{\Nfc}\left(\bar{j}_1,\bar{j}_2\right)
\left(j_1-j_2\ \right)
\sqrtpw{\left(j_1-j_2+\bar{j}_1-\bar{j}_2+1\right) \
\left(-j_1+j_2+\bar{j}_1-\bar{j}_2+1\right) \
\left(j_1+j_2+\bar{j}_1-\bar{j}_2+2\right) \
\left(j_1+j_2+\bar{j}_1-\bar{j}_2+3\right) \
\left(j_1+j_2+\bar{j}_1-\bar{j}_2+4\right) \
\left(j_1+j_2-\bar{j}_1+\bar{j}_2+1\right) \
\left(-j_1-j_2+\bar{j}_1+\bar{j}_2-1\right)
\left(-j_1-j_2+\bar{j}_1+\ \bar{j}_2\right)
\left(j_1+j_2+\bar{j}_1+\bar{j}_2+3\right) \
\left(j_1+j_2+\bar{j}_1+\bar{j}_2+4\right)}}{2 \sqrtpw{2} \
\sqrtpw{\left(j_1+1\right) \left(2 j_1+3\right) \left(j_2+1\right)
\ \left(2 j_2+3\right)}}
\eed,\bed%
\left(
\begin{array}{@{}c@{\;}c@{}||c@{\;}c@{\;\;}c@{\;}c@{}}
 \bar{j}_1+\frac{1}{2} & \bar{j}_2-\frac{1}{2} & \bar{j}_1 & \
\bar{j}_2 & 1 & 1 \\
 j_1-1 & j_2-1 & j_1 & j_2 & 1 & 1
\end{array}
\right)=\zfrac{\text{\Nfc}\left(\bar{j}_1,\bar{j}_2\right)
\left(j_1-j_2\ \right)
\sqrtpw{\left(j_1-j_2+\bar{j}_1-\bar{j}_2+1\right) \
\left(-j_1+j_2+\bar{j}_1-\bar{j}_2+1\right) \
\left(j_1+j_2+\bar{j}_1-\bar{j}_2+1\right) \
\left(j_1+j_2-\bar{j}_1+\bar{j}_2-2\right) \
\left(j_1+j_2-\bar{j}_1+\bar{j}_2-1\right) \
\left(j_1+j_2-\bar{j}_1+\bar{j}_2\right) \
\left(-j_1-j_2+\bar{j}_1+\bar{j}_2+1\right)
\left(-j_1-j_2+\bar{j}_1+\ \bar{j}_2+2\right)
\left(j_1+j_2+\bar{j}_1+\bar{j}_2+1\right) \
\left(j_1+j_2+\bar{j}_1+\bar{j}_2+2\right)}}{2 \sqrtpw{2}
\sqrtpw{j_1 \ \left(2 j_1-1\right) j_2 \left(2 j_2-1\right)}}
\eed,\bed%
\left(
\begin{array}{@{}c@{\;}c@{}||c@{\;}c@{\;\;}c@{\;}c@{}}
 \bar{j}_1+\frac{1}{2} & \bar{j}_2-\frac{1}{2} & \bar{j}_1 & \
\bar{j}_2 & 1 & 1 \\
 j_1-1 & j_2+1 & j_1 & j_2 & 1 & 1
\end{array}
\right)=\zfrac{\text{\Nfc}\left(\bar{j}_1,\bar{j}_2\right) \
\left(j_1+j_2+1\right)
\sqrtpw{\left(j_1-j_2+\bar{j}_1-\bar{j}_2\right) \
\left(-j_1+j_2+\bar{j}_1-\bar{j}_2+1\right)
\left(-j_1+j_2+\bar{j}_1-\ \bar{j}_2+2\right)
\left(-j_1+j_2+\bar{j}_1-\bar{j}_2+3\right) \
\left(j_1+j_2+\bar{j}_1-\bar{j}_2+2\right) \
\left(j_1+j_2-\bar{j}_1+\bar{j}_2\right) \
\left(j_1-j_2+\bar{j}_1+\bar{j}_2\right) \
\left(j_1-j_2+\bar{j}_1+\bar{j}_2+1\right) \
\left(-j_1+j_2+\bar{j}_1+\bar{j}_2+2\right)
\left(-j_1+j_2+\bar{j}_1+\ \bar{j}_2+3\right)}}{2 \sqrtpw{2}
\sqrtpw{j_1 \left(2 j_1-1\right) \ \left(2 j_2^2+5 j_2+3\right)}}
\eed,\bed%
\left(
\begin{array}{@{}c@{\;}c@{}||c@{\;}c@{\;\;}c@{\;}c@{}}
 \bar{j}_1+\frac{1}{2} & \bar{j}_2-\frac{1}{2} & \bar{j}_1 & \
\bar{j}_2 & 1 & 1 \\
 j_1+1 & j_2-1 & j_1 & j_2 & 1 & 1
\end{array}
\right)=-\zfrac{\text{\Nfc}\left(\bar{j}_1,\bar{j}_2\right) \
\left(j_1+j_2+1\right)
\sqrtpw{\left(j_1-j_2+\bar{j}_1-\bar{j}_2+1\right)
\left(j_1-j_2+\bar{j}_1-\bar{j}_2+2\right) \
\left(j_1-j_2+\bar{j}_1-\bar{j}_2+3\right) \
\left(-j_1+j_2+\bar{j}_1-\bar{j}_2\right) \
\left(j_1+j_2+\bar{j}_1-\bar{j}_2+2\right) \
\left(j_1+j_2-\bar{j}_1+\bar{j}_2\right) \
\left(j_1-j_2+\bar{j}_1+\bar{j}_2+2\right) \
\left(j_1-j_2+\bar{j}_1+\bar{j}_2+3\right) \
\left(-j_1+j_2+\bar{j}_1+\bar{j}_2\right) \
\left(-j_1+j_2+\bar{j}_1+\bar{j}_2+1\right)}}{2 \sqrtpw{2} \
\sqrtpw{\left(2 j_1^2+5 j_1+3\right) j_2 \left(2 j_2-1\right)}}
\eed,\bed%
\left(
\begin{array}{@{}c@{\;}c@{}||c@{\;}c@{\;\;}c@{\;}c@{}}
 \bar{j}_1+\frac{1}{2} & \bar{j}_2-\frac{1}{2} & \bar{j}_1 & \
\bar{j}_2 & 1 & 1 \\
 j_1+1 & j_2 & j_1 & j_2 & 1 & 1
\end{array}
\right)=\zfrac{\text{\Nfc}\left(\bar{j}_1,\bar{j}_2\right) \
\sqrtpw{\left(j_1-j_2+\bar{j}_1-\bar{j}_2+1\right) \
\left(j_1-j_2+\bar{j}_1-\bar{j}_2+2\right) \
\left(j_1+j_2+\bar{j}_1-\bar{j}_2+2\right) \
\left(j_1+j_2+\bar{j}_1-\bar{j}_2+3\right) \
\left(-j_1-j_2+\bar{j}_1+\bar{j}_2\right) \
\left(j_1-j_2+\bar{j}_1+\bar{j}_2+2\right) \
\left(-j_1+j_2+\bar{j}_1+\bar{j}_2+1\right) \
\left(j_1+j_2+\bar{j}_1+\bar{j}_2+3\right)} \left(j_2
\left(j_2+1\right)-\left(j_1+1\right)
\left(j_1-\bar{j}_1+\bar{j}_2\right)\right)}{2 \sqrtpw{2}
\sqrtpw{\left(j_1+1\right) \left(2 j_1+3\right) j_2 \
\left(j_2+1\right)}}
\eed,\bed%
\left(
\begin{array}{@{}c@{\;}c@{}||c@{\;}c@{\;\;}c@{\;}c@{}}
 \bar{j}_1+\frac{1}{2} & \bar{j}_2-\frac{1}{2} & \bar{j}_1 & \
\bar{j}_2 & 1 & 1 \\
 j_1-1 & j_2 & j_1 & j_2 & 1 & 1
\end{array}
\right)=-\zfrac{\text{\Nfc}\left(\bar{j}_1,\bar{j}_2\right)
\left(j_1^2+\ \left(\bar{j}_1-\bar{j}_2+1\right) j_1-j_2
\left(j_2+1\right)\right) \
\sqrtpw{\left(-j_1+j_2+\bar{j}_1-\bar{j}_2+1\right) \
\left(-j_1+j_2+\bar{j}_1-\bar{j}_2+2\right) \
\left(j_1+j_2-\bar{j}_1+\bar{j}_2-1\right) \
\left(j_1+j_2-\bar{j}_1+\bar{j}_2\right) \
\left(-j_1-j_2+\bar{j}_1+\bar{j}_2+1\right) \
\left(j_1-j_2+\bar{j}_1+\bar{j}_2+1\right) \
\left(-j_1+j_2+\bar{j}_1+\bar{j}_2+2\right) \
\left(j_1+j_2+\bar{j}_1+\bar{j}_2+2\right)}}{2 \sqrtpw{2}
\sqrtpw{j_1 \ \left(2 j_1-1\right) j_2 \left(j_2+1\right)}}
\eed,\bed%
\left(
\begin{array}{@{}c@{\;}c@{}||c@{\;}c@{\;\;}c@{\;}c@{}}
 \bar{j}_1+\frac{1}{2} & \bar{j}_2-\frac{1}{2} & \bar{j}_1 & \
\bar{j}_2 & 1 & 1 \\
 j_1 & j_2+1 & j_1 & j_2 & 1 & 1
\end{array}
\right)=-\zfrac{\text{\Nfc}\left(\bar{j}_1,\bar{j}_2\right) \
\sqrtpw{\left(-j_1+j_2+\bar{j}_1-\bar{j}_2+1\right) \
\left(-j_1+j_2+\bar{j}_1-\bar{j}_2+2\right) \
\left(j_1+j_2+\bar{j}_1-\bar{j}_2+2\right) \
\left(j_1+j_2+\bar{j}_1-\bar{j}_2+3\right) \
\left(-j_1-j_2+\bar{j}_1+\bar{j}_2\right) \
\left(j_1-j_2+\bar{j}_1+\bar{j}_2+1\right) \
\left(-j_1+j_2+\bar{j}_1+\bar{j}_2+2\right) \
\left(j_1+j_2+\bar{j}_1+\bar{j}_2+3\right)} \
\left(j_1^2+j_1-\left(j_2+1\right)
\left(j_2-\bar{j}_1+\bar{j}_2\right)\right)}{2 \sqrtpw{2}
\sqrtpw{j_1 \left(j_1+1\right) \left(j_2+1\right) \left(2
j_2+3\right)}}
\eed,\bed%
\left(
\begin{array}{@{}c@{\;}c@{}||c@{\;}c@{\;\;}c@{\;}c@{}}
 \bar{j}_1+\frac{1}{2} & \bar{j}_2-\frac{1}{2} & \bar{j}_1 & \
\bar{j}_2 & 1 & 1 \\
 j_1 & j_2-1 & j_1 & j_2 & 1 & 1
\end{array}
\right)=-\zfrac{\text{\Nfc}\left(\bar{j}_1,\bar{j}_2\right) \
\left(j_1^2+j_1-j_2 \left(j_2+\bar{j}_1-\bar{j}_2+1\right)\right)
\ \sqrtpw{\left(j_1-j_2+\bar{j}_1-\bar{j}_2+1\right) \
\left(j_1-j_2+\bar{j}_1-\bar{j}_2+2\right) \
\left(j_1+j_2-\bar{j}_1+\bar{j}_2-1\right) \
\left(j_1+j_2-\bar{j}_1+\bar{j}_2\right) \
\left(-j_1-j_2+\bar{j}_1+\bar{j}_2+1\right) \
\left(j_1-j_2+\bar{j}_1+\bar{j}_2+2\right) \
\left(-j_1+j_2+\bar{j}_1+\bar{j}_2+1\right) \
\left(j_1+j_2+\bar{j}_1+\bar{j}_2+2\right)}}{2 \sqrtpw{2}
\sqrtpw{j_1 \ \left(j_1+1\right) j_2 \left(2 j_2-1\right)}}
\eed,\bed%
\left(
\begin{array}{@{}c@{\;}c@{}||c@{\;}c@{\;\;}c@{\;}c@{}}
 \bar{j}_1+\frac{1}{2} & \bar{j}_2-\frac{1}{2} & \bar{j}_1 & \
\bar{j}_2 & 1 & 1 \\
 j_1 & j_2 & j_1 & j_2 & 1 & 1
\end{array}
\right)=\zfrac{\text{\Nfc}\left(\bar{j}_1,\bar{j}_2\right) \
\left(j_1^2+j_1-j_2 \left(j_2+1\right)\right) \
\sqrtpw{\left(j_1-j_2+\bar{j}_1-\bar{j}_2+1\right) \
\left(-j_1+j_2+\bar{j}_1-\bar{j}_2+1\right) \
\left(j_1+j_2+\bar{j}_1-\bar{j}_2+2\right) \
\left(j_1+j_2-\bar{j}_1+\bar{j}_2\right)} \
\left(-j_1^2-j_1-j_2^2+\left(\bar{j}_1+\bar{j}_2\right){}^2-j_2+3
\ \left(\bar{j}_1+\bar{j}_2\right)+2\right)}{2 \sqrtpw{2}
\sqrtpw{j_1 \ \left(j_1+1\right) j_2 \left(j_2+1\right)}}
\eed,\bed%
\left(
\begin{array}{@{}c@{\;}c@{}||c@{\;}c@{\;\;}c@{\;}c@{}}
 \bar{j}_1+\frac{1}{2} & \bar{j}_2-\frac{1}{2} & \bar{j}_1 & \
\bar{j}_2 & 1 & 1 \\
 j_1+\frac{1}{2} & j_2+\frac{1}{2} & j_1 & j_2 & \frac{1}{2} & \
\frac{1}{2}
\end{array}
\right)=-\zfrac{\text{\Nfc}\left(\bar{j}_1,\bar{j}_2\right) \
\left(j_1-j_2\right) \left(2 j_1+2
j_2-\bar{j}_1+\bar{j}_2+2\right) \
\sqrtpw{\left(j_1-j_2+\bar{j}_1-\bar{j}_2+1\right) \
\left(-j_1+j_2+\bar{j}_1-\bar{j}_2+1\right) \
\left(j_1+j_2+\bar{j}_1-\bar{j}_2+2\right) \
\left(j_1+j_2+\bar{j}_1-\bar{j}_2+3\right) \
\left(-j_1-j_2+\bar{j}_1+\bar{j}_2\right) \
\left(j_1+j_2+\bar{j}_1+\bar{j}_2+3\right)}}{2 \sqrtpw{2} \
\sqrtpw{\left(j_1+1\right) \left(j_2+1\right)}}
\eed,\bed%
\left(
\begin{array}{@{}c@{\;}c@{}||c@{\;}c@{\;\;}c@{\;}c@{}}
 \bar{j}_1+\frac{1}{2} & \bar{j}_2-\frac{1}{2} & \bar{j}_1 & \
\bar{j}_2 & 1 & 1 \\
 j_1-\frac{1}{2} & j_2-\frac{1}{2} & j_1 & j_2 & \frac{1}{2} & \
\frac{1}{2}
\end{array}
\right)=-\zfrac{\text{\Nfc}\left(\bar{j}_1,\bar{j}_2\right) \
\left(j_1-j_2\right) \left(2 j_1+2
j_2+\bar{j}_1-\bar{j}_2+2\right) \
\sqrtpw{\left(j_1-j_2+\bar{j}_1-\bar{j}_2+1\right) \
\left(-j_1+j_2+\bar{j}_1-\bar{j}_2+1\right) \
\left(j_1+j_2-\bar{j}_1+\bar{j}_2-1\right) \
\left(j_1+j_2-\bar{j}_1+\bar{j}_2\right) \
\left(-j_1-j_2+\bar{j}_1+\bar{j}_2+1\right) \
\left(j_1+j_2+\bar{j}_1+\bar{j}_2+2\right)}}{2 \sqrtpw{2}
\sqrtpw{j_1 \ j_2}}
\eed,\bed%
\left(
\begin{array}{@{}c@{\;}c@{}||c@{\;}c@{\;\;}c@{\;}c@{}}
 \bar{j}_1+\frac{1}{2} & \bar{j}_2-\frac{1}{2} & \bar{j}_1 & \
\bar{j}_2 & 1 & 1 \\
 j_1+\frac{1}{2} & j_2-\frac{1}{2} & j_1 & j_2 & \frac{1}{2} & \
\frac{1}{2}
\end{array}
\right)=\zfrac{\text{\Nfc}\left(\bar{j}_1,\bar{j}_2\right) \
\left(j_1+j_2+1\right) \left(2 j_1-2
j_2-\bar{j}_1+\bar{j}_2\right) \
\sqrtpw{\left(j_1-j_2+\bar{j}_1-\bar{j}_2+1\right) \
\left(j_1-j_2+\bar{j}_1-\bar{j}_2+2\right) \
\left(j_1+j_2+\bar{j}_1-\bar{j}_2+2\right) \
\left(j_1+j_2-\bar{j}_1+\bar{j}_2\right) \
\left(j_1-j_2+\bar{j}_1+\bar{j}_2+2\right) \
\left(-j_1+j_2+\bar{j}_1+\bar{j}_2+1\right)}}{2 \sqrtpw{2} \
\sqrtpw{\left(j_1+1\right) j_2}}
\eed,\bed%
\left(
\begin{array}{@{}c@{\;}c@{}||c@{\;}c@{\;\;}c@{\;}c@{}}
 \bar{j}_1+\frac{1}{2} & \bar{j}_2-\frac{1}{2} & \bar{j}_1 & \
\bar{j}_2 & 1 & 1 \\
 j_1-\frac{1}{2} & j_2+\frac{1}{2} & j_1 & j_2 & \frac{1}{2} & \
\frac{1}{2}
\end{array}
\right)=\zfrac{\text{\Nfc}\left(\bar{j}_1,\bar{j}_2\right) \
\left(j_1+j_2+1\right) \left(2 j_1-2
j_2+\bar{j}_1-\bar{j}_2\right) \
\sqrtpw{\left(-j_1+j_2+\bar{j}_1-\bar{j}_2+1\right) \
\left(-j_1+j_2+\bar{j}_1-\bar{j}_2+2\right) \
\left(j_1+j_2+\bar{j}_1-\bar{j}_2+2\right) \
\left(j_1+j_2-\bar{j}_1+\bar{j}_2\right) \
\left(j_1-j_2+\bar{j}_1+\bar{j}_2+1\right) \
\left(-j_1+j_2+\bar{j}_1+\bar{j}_2+2\right)}}{2 \sqrtpw{2}
\sqrtpw{j_1 \ \left(j_2+1\right)}}
\eed,\bed%
\left(
\begin{array}{@{}c@{\;}c@{}||c@{\;}c@{\;\;}c@{\;}c@{}}
 \bar{j}_1+\frac{1}{2} & \bar{j}_2-\frac{1}{2} & \bar{j}_1 & \
\bar{j}_2 & 1 & 1 \\
 j_1 & j_2 & j_1 & j_2 & 0 & 0
\end{array}
\right)=\sqrtpw{\frac{5}{2}}
\text{\Nfc}\left(\bar{j}_1,\bar{j}_2\right) \ \left(j_1-j_2\right)
\left(j_1+j_2+1\right) \
\sqrtpw{\left(j_1-j_2+\bar{j}_1-\bar{j}_2+1\right) \
\left(-j_1+j_2+\bar{j}_1-\bar{j}_2+1\right) \
\left(j_1+j_2+\bar{j}_1-\bar{j}_2+2\right) \
\left(j_1+j_2-\bar{j}_1+\bar{j}_2\right)} \eed%
.

\vspace{1cm} The $(\bar{j}_1, \bar{j}_2)$ representation appears twice in
decomposition of $(\bar{j}_1, \bar{j}_2) \otimes (\bar 1, \bar 1)$.
In order to distinguish them, we follow Wong's convention. The CG
coefficients for $(\bar{j}_1, \bar{j}_2) \otimes (\bar 1, \bar 1) \rightarrow
(\bar{j}_1, \bar{j}_2)_1$ are:

\bed \text{\Ngc}\left(\bar{j}_1,\bar{j}_2\right) =2 \sqrtpw{5}
\left(4 \bar{j}_2^2 \left(\bar{j}_2+1\right){}^2+11 \left(8
\bar{j}_1^2+16 \bar{j}_1+5\right) \bar{j}_2
   \left(\bar{j}_2+1\right)+\bar{j}_1 \left(\bar{j}_1+2\right) \left(2
     \bar{j}_1-1\right) \left(2 \bar{j}_1+5\right)\right)^{-\frac 12} \eed

\bed \left(
\begin{array}{@{}c@{\;}c@{}||c@{\;}c@{\;\;}c@{\;}c@{}}
 \bar{j}_1 & \bar{j}_2 & \bar{j}_1 & \bar{j}_2 & 1 & 1 \\
 j_1+1 & j_2+1 & j_1 & j_2 & 1 & 1
\end{array}
\right)_1=-\zfrac{\text{\Ngc}\left(\bar{j}_1,\bar{j}_2\right) \
\sqrtpw{\left(j_1+j_2-\bar{j}_1-\bar{j}_2\right) \
\left(j_1+j_2-\bar{j}_1-\bar{j}_2+1\right) \
\left(j_1+j_2+\bar{j}_1-\bar{j}_2+2\right) \
\left(j_1+j_2+\bar{j}_1-\bar{j}_2+3\right) \
\left(j_1+j_2-\bar{j}_1+\bar{j}_2+1\right) \
\left(j_1+j_2-\bar{j}_1+\bar{j}_2+2\right) \
\left(j_1+j_2+\bar{j}_1+\bar{j}_2+3\right) \
\left(j_1+j_2+\bar{j}_1+\bar{j}_2+4\right)}}{8
\sqrtpw{\left(j_1+1\right) \left(2 j_1+3\right) \left(j_2+1\right)
\left(2 j_2+3\right)}}
\eed,\bed%
\left(
\begin{array}{@{}c@{\;}c@{}||c@{\;}c@{\;\;}c@{\;}c@{}}
 \bar{j}_1 & \bar{j}_2 & \bar{j}_1 & \bar{j}_2 & 1 & 1 \\
 j_1-1 & j_2-1 & j_1 & j_2 & 1 & 1
\end{array}
\right)_1=-\zfrac{\text{\Ngc}\left(\bar{j}_1,\bar{j}_2\right) \
\sqrtpw{\left(j_1+j_2+\bar{j}_1-\bar{j}_2\right) \
\left(j_1+j_2+\bar{j}_1-\bar{j}_2+1\right) \
\left(j_1+j_2-\bar{j}_1+\bar{j}_2-1\right) \
\left(j_1+j_2-\bar{j}_1+\bar{j}_2\right) \
\left(-j_1-j_2+\bar{j}_1+\bar{j}_2+1\right)
\left(-j_1-j_2+\bar{j}_1+\ \bar{j}_2+2\right)
\left(j_1+j_2+\bar{j}_1+\bar{j}_2+1\right) \
\left(j_1+j_2+\bar{j}_1+\bar{j}_2+2\right)}}{8 \sqrtpw{j_1 \left(2
\ j_1-1\right) j_2 \left(2 j_2-1\right)}}
\eed,\bed%
\left(
\begin{array}{@{}c@{\;}c@{}||c@{\;}c@{\;\;}c@{\;}c@{}}
 \bar{j}_1 & \bar{j}_2 & \bar{j}_1 & \bar{j}_2 & 1 & 1 \\
 j_1-1 & j_2+1 & j_1 & j_2 & 1 & 1
\end{array}
\right)_1=-\zfrac{\text{\Ngc}\left(\bar{j}_1,\bar{j}_2\right) \
\sqrtpw{\left(j_1-j_2+\bar{j}_1-\bar{j}_2-1\right) \
\left(j_1-j_2+\bar{j}_1-\bar{j}_2\right) \
\left(-j_1+j_2+\bar{j}_1-\bar{j}_2+1\right)
\left(-j_1+j_2+\bar{j}_1-\ \bar{j}_2+2\right)
\left(j_1-j_2+\bar{j}_1+\bar{j}_2\right) \
\left(j_1-j_2+\bar{j}_1+\bar{j}_2+1\right) \
\left(-j_1+j_2+\bar{j}_1+\bar{j}_2+2\right)
\left(-j_1+j_2+\bar{j}_1+\ \bar{j}_2+3\right)}}{8 \sqrtpw{j_1
\left(2 j_1-1\right) \left(2 j_2^2+5 \ j_2+3\right)}}
\eed,\bed%
\left(
\begin{array}{@{}c@{\;}c@{}||c@{\;}c@{\;\;}c@{\;}c@{}}
 \bar{j}_1 & \bar{j}_2 & \bar{j}_1 & \bar{j}_2 & 1 & 1 \\
 j_1+1 & j_2-1 & j_1 & j_2 & 1 & 1
\end{array}
\right)_1=-\zfrac{\text{\Ngc}\left(\bar{j}_1,\bar{j}_2\right) \
\sqrtpw{\left(j_1-j_2+\bar{j}_1-\bar{j}_2+1\right) \
\left(j_1-j_2+\bar{j}_1-\bar{j}_2+2\right) \
\left(-j_1+j_2+\bar{j}_1-\bar{j}_2-1\right)
\left(-j_1+j_2+\bar{j}_1-\ \bar{j}_2\right)
\left(j_1-j_2+\bar{j}_1+\bar{j}_2+2\right) \
\left(j_1-j_2+\bar{j}_1+\bar{j}_2+3\right) \
\left(-j_1+j_2+\bar{j}_1+\bar{j}_2\right) \
\left(-j_1+j_2+\bar{j}_1+\bar{j}_2+1\right)}}{8 \sqrtpw{\left(2
j_1^2+5 \ j_1+3\right) j_2 \left(2 j_2-1\right)}}
\eed,\bed%
\left(
\begin{array}{@{}c@{\;}c@{}||c@{\;}c@{\;\;}c@{\;}c@{}}
 \bar{j}_1 & \bar{j}_2 & \bar{j}_1 & \bar{j}_2 & 1 & 1 \\
 j_1+1 & j_2 & j_1 & j_2 & 1 & 1
\end{array}
\right)_1=-\zfrac{\text{\Ngc}\left(\bar{j}_1,\bar{j}_2\right) \
\sqrtpw{\left(j_1-j_2+\bar{j}_1-\bar{j}_2+1\right) \
\left(-j_1+j_2+\bar{j}_1-\bar{j}_2\right) \
\left(j_1+j_2+\bar{j}_1-\bar{j}_2+2\right) \
\left(j_1+j_2-\bar{j}_1+\bar{j}_2+1\right) \
\left(-j_1-j_2+\bar{j}_1+\bar{j}_2\right) \
\left(j_1-j_2+\bar{j}_1+\bar{j}_2+2\right) \
\left(-j_1+j_2+\bar{j}_1+\bar{j}_2+1\right) \
\left(j_1+j_2+\bar{j}_1+\bar{j}_2+3\right)}}{8
\sqrtpw{\left(j_1+1\right) \left(2 j_1+3\right) j_2
\left(j_2+1\right)}}
\eed,\bed%
\left(
\begin{array}{@{}c@{\;}c@{}||c@{\;}c@{\;\;}c@{\;}c@{}}
 \bar{j}_1 & \bar{j}_2 & \bar{j}_1 & \bar{j}_2 & 1 & 1 \\
 j_1-1 & j_2 & j_1 & j_2 & 1 & 1
\end{array}
\right)_1=\zfrac{\text{\Ngc}\left(\bar{j}_1,\bar{j}_2\right) \
\sqrtpw{\left(j_1-j_2+\bar{j}_1-\bar{j}_2\right) \
\left(-j_1+j_2+\bar{j}_1-\bar{j}_2+1\right) \
\left(j_1+j_2+\bar{j}_1-\bar{j}_2+1\right) \
\left(j_1+j_2-\bar{j}_1+\bar{j}_2\right) \
\left(-j_1-j_2+\bar{j}_1+\bar{j}_2+1\right) \
\left(j_1-j_2+\bar{j}_1+\bar{j}_2+1\right) \
\left(-j_1+j_2+\bar{j}_1+\bar{j}_2+2\right) \
\left(j_1+j_2+\bar{j}_1+\bar{j}_2+2\right)}}{8 \sqrtpw{j_1 \left(2
\ j_1-1\right) j_2 \left(j_2+1\right)}}
\eed,\bed%
\left(
\begin{array}{@{}c@{\;}c@{}||c@{\;}c@{\;\;}c@{\;}c@{}}
 \bar{j}_1 & \bar{j}_2 & \bar{j}_1 & \bar{j}_2 & 1 & 1 \\
 j_1 & j_2+1 & j_1 & j_2 & 1 & 1
\end{array}
\right)_1=-\zfrac{\text{\Ngc}\left(\bar{j}_1,\bar{j}_2\right) \
\sqrtpw{\left(j_1-j_2+\bar{j}_1-\bar{j}_2\right) \
\left(-j_1+j_2+\bar{j}_1-\bar{j}_2+1\right) \
\left(j_1+j_2+\bar{j}_1-\bar{j}_2+2\right) \
\left(j_1+j_2-\bar{j}_1+\bar{j}_2+1\right) \
\left(-j_1-j_2+\bar{j}_1+\bar{j}_2\right) \
\left(j_1-j_2+\bar{j}_1+\bar{j}_2+1\right) \
\left(-j_1+j_2+\bar{j}_1+\bar{j}_2+2\right) \
\left(j_1+j_2+\bar{j}_1+\bar{j}_2+3\right)}}{8 \sqrtpw{j_1
\left(j_1+1\right) \left(j_2+1\right) \left(2 j_2+3\right)}}
\eed,\bed%
\left(
\begin{array}{@{}c@{\;}c@{}||c@{\;}c@{\;\;}c@{\;}c@{}}
 \bar{j}_1 & \bar{j}_2 & \bar{j}_1 & \bar{j}_2 & 1 & 1 \\
 j_1 & j_2-1 & j_1 & j_2 & 1 & 1
\end{array}
\right)_1=\zfrac{\text{\Ngc}\left(\bar{j}_1,\bar{j}_2\right) \
\sqrtpw{\left(j_1-j_2+\bar{j}_1-\bar{j}_2+1\right) \
\left(-j_1+j_2+\bar{j}_1-\bar{j}_2\right) \
\left(j_1+j_2+\bar{j}_1-\bar{j}_2+1\right) \
\left(j_1+j_2-\bar{j}_1+\bar{j}_2\right) \
\left(-j_1-j_2+\bar{j}_1+\bar{j}_2+1\right) \
\left(j_1-j_2+\bar{j}_1+\bar{j}_2+2\right) \
\left(-j_1+j_2+\bar{j}_1+\bar{j}_2+1\right) \
\left(j_1+j_2+\bar{j}_1+\bar{j}_2+2\right)}}{8 \sqrtpw{j_1
\left(j_1+1\right) j_2 \left(2 j_2-1\right)}}
\eed,\bed%
\left(
\begin{array}{@{}c@{\;}c@{}||c@{\;}c@{\;\;}c@{\;}c@{}}
 \bar{j}_1 & \bar{j}_2 & \bar{j}_1 & \bar{j}_2 & 1 & 1 \\
 j_1 & j_2 & j_1 & j_2 & 1 & 1
\end{array}
\right)_1=-\zfrac{\text{\Ngc}\left(\bar{j}_1,\bar{j}_2\right) \
\left(j_1^4+2 j_1^3-\left(10 j_2^2+10 j_2+2 \bar{j}_1^2+2 \
\bar{j}_2^2+4 \bar{j}_1+2 \bar{j}_2+1\right) j_1^2-2 \left(5
j_2^2+5 \ j_2+\bar{j}_1^2+\bar{j}_2^2+2
\bar{j}_1+\bar{j}_2+1\right) j_1+j_2^4+\ \bar{j}_1^4+\bar{j}_2^4+2
j_2^3+4 \bar{j}_1^3+2 \bar{j}_2^3+5 \ \bar{j}_1^2-2 \bar{j}_1^2
\bar{j}_2^2-4 \bar{j}_1 \ \bar{j}_2^2-\bar{j}_2^2+2 \bar{j}_1-2
\bar{j}_1^2 \bar{j}_2-4 \ \bar{j}_1 \bar{j}_2-2 \bar{j}_2-2 j_2
\left(\bar{j}_1^2+2 \
\bar{j}_1+\bar{j}_2^2+\bar{j}_2+1\right)-j_2^2 \left(2
\bar{j}_1^2+4 \ \bar{j}_1+2 \bar{j}_2^2+2
\bar{j}_2+1\right)\right)}{8 \sqrtpw{j_1 \ \left(j_1+1\right) j_2
\left(j_2+1\right)}}
\eed,\bed%
\left(
\begin{array}{@{}c@{\;}c@{}||c@{\;}c@{\;\;}c@{\;}c@{}}
 \bar{j}_1 & \bar{j}_2 & \bar{j}_1 & \bar{j}_2 & 1 & 1 \\
 j_1+\frac{1}{2} & j_2+\frac{1}{2} & j_1 & j_2 & \frac{1}{2} & \
\frac{1}{2}
\end{array}
\right)_1=\zfrac{\text{\Ngc}\left(\bar{j}_1,\bar{j}_2\right)
\left(2 j_1+2 \ j_2+3\right)
\sqrtpw{\left(j_1+j_2+\bar{j}_1-\bar{j}_2+2\right) \
\left(j_1+j_2-\bar{j}_1+\bar{j}_2+1\right) \
\left(-j_1-j_2+\bar{j}_1+\bar{j}_2\right) \
\left(j_1+j_2+\bar{j}_1+\bar{j}_2+3\right)}}{8
\sqrtpw{\left(j_1+1\right) \left(j_2+1\right)}}
\eed,\bed%
\left(
\begin{array}{@{}c@{\;}c@{}||c@{\;}c@{\;\;}c@{\;}c@{}}
 \bar{j}_1 & \bar{j}_2 & \bar{j}_1 & \bar{j}_2 & 1 & 1 \\
 j_1-\frac{1}{2} & j_2-\frac{1}{2} & j_1 & j_2 & \frac{1}{2} & \
\frac{1}{2}
\end{array}
\right)_1=\zfrac{\text{\Ngc}\left(\bar{j}_1,\bar{j}_2\right)
\left(2 j_1+2 \ j_2+1\right)
\sqrtpw{\left(j_1+j_2+\bar{j}_1-\bar{j}_2+1\right) \
\left(j_1+j_2-\bar{j}_1+\bar{j}_2\right) \
\left(-j_1-j_2+\bar{j}_1+\bar{j}_2+1\right) \
\left(j_1+j_2+\bar{j}_1+\bar{j}_2+2\right)}}{8 \sqrtpw{j_1 j_2}}
\eed,\bed%
\left(
\begin{array}{@{}c@{\;}c@{}||c@{\;}c@{\;\;}c@{\;}c@{}}
 \bar{j}_1 & \bar{j}_2 & \bar{j}_1 & \bar{j}_2 & 1 & 1 \\
 j_1+\frac{1}{2} & j_2-\frac{1}{2} & j_1 & j_2 & \frac{1}{2} & \
\frac{1}{2}
\end{array}
\right)_1=\zfrac{\text{\Ngc}\left(\bar{j}_1,\bar{j}_2\right)
\left(2 j_1-2 \ j_2+1\right)
\sqrtpw{\left(j_1-j_2+\bar{j}_1-\bar{j}_2+1\right) \
\left(-j_1+j_2+\bar{j}_1-\bar{j}_2\right) \
\left(j_1-j_2+\bar{j}_1+\bar{j}_2+2\right) \
\left(-j_1+j_2+\bar{j}_1+\bar{j}_2+1\right)}}{8
\sqrtpw{\left(j_1+1\right) j_2}}
\eed,\bed%
\left(
\begin{array}{@{}c@{\;}c@{}||c@{\;}c@{\;\;}c@{\;}c@{}}
 \bar{j}_1 & \bar{j}_2 & \bar{j}_1 & \bar{j}_2 & 1 & 1 \\
 j_1-\frac{1}{2} & j_2+\frac{1}{2} & j_1 & j_2 & \frac{1}{2} & \
\frac{1}{2}
\end{array}
\right)_1=-\zfrac{\text{\Ngc}\left(\bar{j}_1,\bar{j}_2\right)
\left(2 \ j_1-2 j_2-1\right)
\sqrtpw{\left(j_1-j_2+\bar{j}_1-\bar{j}_2\right) \
\left(-j_1+j_2+\bar{j}_1-\bar{j}_2+1\right) \
\left(j_1-j_2+\bar{j}_1+\bar{j}_2+1\right) \
\left(-j_1+j_2+\bar{j}_1+\bar{j}_2+2\right)}}{8 \sqrtpw{j_1
\left(j_2+1\ \right)}}
\eed,\bed%
\left(
\begin{array}{@{}c@{\;}c@{}||c@{\;}c@{\;\;}c@{\;}c@{}}
 \bar{j}_1 & \bar{j}_2 & \bar{j}_1 & \bar{j}_2 & 1 & 1 \\
 j_1 & j_2 & j_1 & j_2 & 0 & 0
\end{array}
\right)_1=-\zfrac{\text{\Ngc}\left(\bar{j}_1,\bar{j}_2\right)
\left(5 \ j_1^2+5 j_1+5 j_2^2+5 j_2-3 \left(\bar{j}_1^2+2 \
\bar{j}_1+\bar{j}_2^2+\bar{j}_2\right)\right)}{2 \sqrtpw{5}} \eed%
.

\vspace{1cm}
The CG coefficients for $(\bar{j}_1, \bar{j}_2)
\otimes (\bar 1, \bar 1) \rightarrow (\bar{j}_1, \bar{j}_2)_2$
read:

\bed%
\left(
\begin{array}{@{}c@{\;}c@{}||c@{\;}c@{\;\;}c@{\;}c@{}}
 \bar{j}_1 & \bar{j}_2 & \bar{j}_1 & \bar{j}_2 & 1 & 1 \\
 j'_1 & j'_2 & j_1 & j_2 & J_1 & J_2
\end{array}
\right)_2=  \sqrtpw{H^2 - X^2}\left(
\begin{array}{@{}c@{\;}c@{}||c@{\;}c@{\;\;}c@{\;}c@{}}
 \bar{j}_1 & \bar{j}_2 & \bar{j}_1 & \bar{j}_2 & 1 & 1 \\
 j'_1 & j'_2 & j_1 & j_2 & J_1 & J_2
\end{array}
\right)_\tmp  - X \left(
\begin{array}{@{}c@{\;}c@{}||c@{\;}c@{\;\;}c@{\;}c@{}}
 \bar{j}_1 & \bar{j}_2 & \bar{j}_1 & \bar{j}_2 & 1 & 1 \\
 j'_1 & j'_2 & j_1 & j_2 & J_1 & J_2
\end{array}
\right)_1,\eed %
where%
\bed X = -\frac{1}{10} \text{\Ngc}\left(\bar{j}_1,\bar{j}_2\right)
\left(\bar{j}_1-\bar{j}_2\right)
\left(\bar{j}_1-\bar{j}_2+1\right)
   \left(\bar{j}_1+\bar{j}_2+1\right) \left(\bar{j}_1+\bar{j}_2+2\right)
   \left(4 \bar{j}_1 \left(\bar{j}_1+2\right)+4 \bar{j}_2
   \left(\bar{j}_2+1\right)-5\right),\eed
\bed H^2 = \frac{1}{5} \left(\bar{j}_1-\bar{j}_2\right)
\left(\bar{j}_1-\bar{j}_2+1\right)
\left(\bar{j}_1+\bar{j}_2+1\right)
   \left(\bar{j}_1+\bar{j}_2+2\right) \left(4 \bar{j}_2^4+8
     \bar{j}_2^3-\left(8 \bar{j}_1 \left(\bar{j}_1+2\right)+9\right)
     \bar{j}_2^2-\left(8
   \bar{j}_1 \left(\bar{j}_1+2\right)+13\right)
 \bar{j}_2+\left(\bar{j}_1+1\right){}^2 \left(4 \bar{j}_1
\left(\bar{j}_1+2\right)-5\right)\right),\;\;\;\;\;\;\;\;\;\;\;\;\;\;\;\;\;\;\;
\eed
and the list of the auxiliary coefficients is as follows:

\bed \left(
\begin{array}{@{}c@{\;}c@{}||c@{\;}c@{\;\;}c@{\;}c@{}}
 \bar{j}_1 & \bar{j}_2 & \bar{j}_1 & \bar{j}_2 & 1 & 1 \\
 j_1+1 & j_2+1 & j_1 & j_2 & 1 & 1
\end{array}
\right)_\tmp=\zfrac{\left(j_1-j_2\right){}^2
\sqrtpw{\left(j_1+j_2-\bar{j}_1-\ \bar{j}_2\right)
\left(j_1+j_2-\bar{j}_1-\bar{j}_2+1\right) \
\left(j_1+j_2+\bar{j}_1-\bar{j}_2+2\right) \
\left(j_1+j_2+\bar{j}_1-\bar{j}_2+3\right) \
\left(j_1+j_2-\bar{j}_1+\bar{j}_2+1\right) \
\left(j_1+j_2-\bar{j}_1+\bar{j}_2+2\right) \
\left(j_1+j_2+\bar{j}_1+\bar{j}_2+3\right) \
\left(j_1+j_2+\bar{j}_1+\bar{j}_2+4\right)}}{4
\sqrtpw{\left(j_1+1\right) \left(2 j_1+3\right) \left(j_2+1\right)
\left(2 j_2+3\right)}}
\eed,\bed%
\left(
\begin{array}{@{}c@{\;}c@{}||c@{\;}c@{\;\;}c@{\;}c@{}}
 \bar{j}_1 & \bar{j}_2 & \bar{j}_1 & \bar{j}_2 & 1 & 1 \\
 j_1-1 & j_2-1 & j_1 & j_2 & 1 & 1
\end{array}
\right)_\tmp=\zfrac{\left(j_1-j_2\right){}^2
\sqrtpw{\left(j_1+j_2+\bar{j}_1-\ \bar{j}_2\right)
\left(j_1+j_2+\bar{j}_1-\bar{j}_2+1\right) \
\left(j_1+j_2-\bar{j}_1+\bar{j}_2-1\right) \
\left(j_1+j_2-\bar{j}_1+\bar{j}_2\right) \
\left(-j_1-j_2+\bar{j}_1+\bar{j}_2+1\right)
\left(-j_1-j_2+\bar{j}_1+\ \bar{j}_2+2\right)
\left(j_1+j_2+\bar{j}_1+\bar{j}_2+1\right) \
\left(j_1+j_2+\bar{j}_1+\bar{j}_2+2\right)}}{4 \sqrtpw{j_1 \left(2
\ j_1-1\right) j_2 \left(2 j_2-1\right)}}
\eed,\bed%
\left(
\begin{array}{@{}c@{\;}c@{}||c@{\;}c@{\;\;}c@{\;}c@{}}
 \bar{j}_1 & \bar{j}_2 & \bar{j}_1 & \bar{j}_2 & 1 & 1 \\
 j_1-1 & j_2+1 & j_1 & j_2 & 1 & 1
\end{array}
\right)_\tmp=\zfrac{\left(j_1+j_2+1\right){}^2 \
\sqrtpw{\left(j_1-j_2+\bar{j}_1-\bar{j}_2-1\right) \
\left(j_1-j_2+\bar{j}_1-\bar{j}_2\right) \
\left(-j_1+j_2+\bar{j}_1-\bar{j}_2+1\right)
\left(-j_1+j_2+\bar{j}_1-\ \bar{j}_2+2\right)
\left(j_1-j_2+\bar{j}_1+\bar{j}_2\right) \
\left(j_1-j_2+\bar{j}_1+\bar{j}_2+1\right) \
\left(-j_1+j_2+\bar{j}_1+\bar{j}_2+2\right)
\left(-j_1+j_2+\bar{j}_1+\ \bar{j}_2+3\right)}}{4 \sqrtpw{j_1
\left(2 j_1-1\right) \left(2 j_2^2+5 \ j_2+3\right)}}
\eed,\bed%
\left(
\begin{array}{@{}c@{\;}c@{}||c@{\;}c@{\;\;}c@{\;}c@{}}
 \bar{j}_1 & \bar{j}_2 & \bar{j}_1 & \bar{j}_2 & 1 & 1 \\
 j_1+1 & j_2-1 & j_1 & j_2 & 1 & 1
\end{array}
\right)_\tmp=\zfrac{\left(j_1+j_2+1\right){}^2 \
\sqrtpw{\left(j_1-j_2+\bar{j}_1-\bar{j}_2+1\right) \
\left(j_1-j_2+\bar{j}_1-\bar{j}_2+2\right) \
\left(-j_1+j_2+\bar{j}_1-\bar{j}_2-1\right)
\left(-j_1+j_2+\bar{j}_1-\ \bar{j}_2\right)
\left(j_1-j_2+\bar{j}_1+\bar{j}_2+2\right) \
\left(j_1-j_2+\bar{j}_1+\bar{j}_2+3\right) \
\left(-j_1+j_2+\bar{j}_1+\bar{j}_2\right) \
\left(-j_1+j_2+\bar{j}_1+\bar{j}_2+1\right)}}{4 \sqrtpw{\left(2
j_1^2+5 \ j_1+3\right) j_2 \left(2 j_2-1\right)}}
\eed,\bed%
\left(
\begin{array}{@{}c@{\;}c@{}||c@{\;}c@{\;\;}c@{\;}c@{}}
 \bar{j}_1 & \bar{j}_2 & \bar{j}_1 & \bar{j}_2 & 1 & 1 \\
 j_1+1 & j_2 & j_1 & j_2 & 1 & 1
\end{array}
\right)_\tmp=\zfrac{\sqrtpw{j_2 \left(j_2+1\right)}
\left(\frac{\left(j_1+1\right){}^2}{j_2
\left(j_2+1\right)}-1\right) \
\sqrtpw{\left(j_1-j_2+\bar{j}_1-\bar{j}_2+1\right) \
\left(-j_1+j_2+\bar{j}_1-\bar{j}_2\right) \
\left(j_1+j_2+\bar{j}_1-\bar{j}_2+2\right) \
\left(j_1+j_2-\bar{j}_1+\bar{j}_2+1\right) \
\left(-j_1-j_2+\bar{j}_1+\bar{j}_2\right) \
\left(j_1-j_2+\bar{j}_1+\bar{j}_2+2\right) \
\left(-j_1+j_2+\bar{j}_1+\bar{j}_2+1\right) \
\left(j_1+j_2+\bar{j}_1+\bar{j}_2+3\right)}}{4
\sqrtpw{\left(j_1+1\right) \left(2 j_1+3\right)}}
\eed,\bed%
\left(
\begin{array}{@{}c@{\;}c@{}||c@{\;}c@{\;\;}c@{\;}c@{}}
 \bar{j}_1 & \bar{j}_2 & \bar{j}_1 & \bar{j}_2 & 1 & 1 \\
 j_1-1 & j_2 & j_1 & j_2 & 1 & 1
\end{array}
\right)_\tmp=\zfrac{\left(-j_1^2+j_2^2+j_2\right) \
\sqrtpw{\left(j_1-j_2+\bar{j}_1-\bar{j}_2\right) \
\left(-j_1+j_2+\bar{j}_1-\bar{j}_2+1\right) \
\left(j_1+j_2+\bar{j}_1-\bar{j}_2+1\right) \
\left(j_1+j_2-\bar{j}_1+\bar{j}_2\right) \
\left(-j_1-j_2+\bar{j}_1+\bar{j}_2+1\right) \
\left(j_1-j_2+\bar{j}_1+\bar{j}_2+1\right) \
\left(-j_1+j_2+\bar{j}_1+\bar{j}_2+2\right) \
\left(j_1+j_2+\bar{j}_1+\bar{j}_2+2\right)}}{4 \sqrtpw{j_1 \left(2
\ j_1-1\right)} \sqrtpw{j_2 \left(j_2+1\right)}}
\eed,\bed%
\left(
\begin{array}{@{}c@{\;}c@{}||c@{\;}c@{\;\;}c@{\;}c@{}}
 \bar{j}_1 & \bar{j}_2 & \bar{j}_1 & \bar{j}_2 & 1 & 1 \\
 j_1 & j_2+1 & j_1 & j_2 & 1 & 1
\end{array}
\right)_\tmp=\zfrac{\sqrtpw{j_1 \left(j_1+1\right)}
\left(\frac{\left(j_2+1\right){}^2}{j_1
\left(j_1+1\right)}-1\right) \
\sqrtpw{\left(j_1-j_2+\bar{j}_1-\bar{j}_2\right) \
\left(-j_1+j_2+\bar{j}_1-\bar{j}_2+1\right) \
\left(j_1+j_2+\bar{j}_1-\bar{j}_2+2\right) \
\left(j_1+j_2-\bar{j}_1+\bar{j}_2+1\right) \
\left(-j_1-j_2+\bar{j}_1+\bar{j}_2\right) \
\left(j_1-j_2+\bar{j}_1+\bar{j}_2+1\right) \
\left(-j_1+j_2+\bar{j}_1+\bar{j}_2+2\right) \
\left(j_1+j_2+\bar{j}_1+\bar{j}_2+3\right)}}{4
\sqrtpw{\left(j_2+1\right) \left(2 j_2+3\right)}}
\eed,\bed%
\left(
\begin{array}{@{}c@{\;}c@{}||c@{\;}c@{\;\;}c@{\;}c@{}}
 \bar{j}_1 & \bar{j}_2 & \bar{j}_1 & \bar{j}_2 & 1 & 1 \\
 j_1 & j_2-1 & j_1 & j_2 & 1 & 1
\end{array}
\right)_\tmp=\zfrac{\left(j_1^2+j_1-j_2^2\right) \
\sqrtpw{\left(j_1-j_2+\bar{j}_1-\bar{j}_2+1\right) \
\left(-j_1+j_2+\bar{j}_1-\bar{j}_2\right) \
\left(j_1+j_2+\bar{j}_1-\bar{j}_2+1\right) \
\left(j_1+j_2-\bar{j}_1+\bar{j}_2\right) \
\left(-j_1-j_2+\bar{j}_1+\bar{j}_2+1\right) \
\left(j_1-j_2+\bar{j}_1+\bar{j}_2+2\right) \
\left(-j_1+j_2+\bar{j}_1+\bar{j}_2+1\right) \
\left(j_1+j_2+\bar{j}_1+\bar{j}_2+2\right)}}{4 \sqrtpw{j_1
\left(j_1+1\right)} \sqrtpw{j_2 \left(2 j_2-1\right)}}
\eed,\bed%
\left(
\begin{array}{@{}c@{\;}c@{}||c@{\;}c@{\;\;}c@{\;}c@{}}
 \bar{j}_1 & \bar{j}_2 & \bar{j}_1 & \bar{j}_2 & 1 & 1 \\
 j_1 & j_2 & j_1 & j_2 & 1 & 1
\end{array}
\right)_\tmp=-\zfrac{j_1^6+3 j_1^5-\left(j_2^2+j_2+2 \bar{j}_1^2+2
\ \bar{j}_2^2+4 \bar{j}_1+2 \bar{j}_2-1\right) j_1^4-\left(2
j_2^2+2 \ j_2+4 \bar{j}_1^2+4 \bar{j}_2^2+8 \bar{j}_1+4
\bar{j}_2+3\right) \ j_1^3+\left(-j_2^4-2 j_2^3+\left(4
\bar{j}_1^2+8 \bar{j}_1+4 \ \bar{j}_2^2+4 \bar{j}_2+2\right)
j_2^2+\left(4 \bar{j}_1^2+8 \ \bar{j}_1+4 \bar{j}_2^2+4
\bar{j}_2+3\right) \ j_2+\bar{j}_1^4+\bar{j}_2^4+4 \bar{j}_1^3+2
\bar{j}_2^3-3 \ \bar{j}_2^2-4 \bar{j}_2+\bar{j}_1^2 \left(-2
\bar{j}_2^2-2 \ \bar{j}_2+3\right)-2 \bar{j}_1 \left(2
\bar{j}_2^2+2 \bar{j}_2+1\right)-2\right) j_1^2+\left(-j_2^4-2
j_2^3+\left(4 \bar{j}_1^2+8 \ \bar{j}_1+4 \bar{j}_2^2+4
\bar{j}_2+3\right) j_2^2+4 \ \left(\bar{j}_1^2+2
\bar{j}_1+\bar{j}_2^2+\bar{j}_2+1\right) \ j_2+\bar{j}_1^4+4
\bar{j}_1^3+\bar{j}_1 \left(-4 \bar{j}_2^2-4 \
\bar{j}_2+2\right)+\bar{j}_1^2 \left(-2 \bar{j}_2^2-2
\bar{j}_2+5\right)+\bar{j}_2 \left(\bar{j}_2^3+2
\bar{j}_2^2-\bar{j}_2-2\right)\right) j_1+j_2 \left(j_2+1\right)
\left(j_2^4+2 j_2^3-\left(2 \ \bar{j}_1^2+4 \bar{j}_1+2
\bar{j}_2^2+2 \bar{j}_2+1\right) j_2^2-2 \ \left(\bar{j}_1^2+2
\bar{j}_1+\bar{j}_2^2+\bar{j}_2+1\right) \ j_2+\bar{j}_1^4+4
\bar{j}_1^3+\bar{j}_1 \left(-4 \bar{j}_2^2-4 \
\bar{j}_2+2\right)+\bar{j}_1^2 \left(-2 \bar{j}_2^2-2
\bar{j}_2+5\right)+\bar{j}_2 \left(\bar{j}_2^3+2
\bar{j}_2^2-\bar{j}_2-2\right)\right)}{4 \sqrtpw{j_1
\left(j_1+1\right) j_2 \left(j_2+1\right)}}
\eed,\bed%
\left(
\begin{array}{@{}c@{\;}c@{}||c@{\;}c@{\;\;}c@{\;}c@{}}
 \bar{j}_1 & \bar{j}_2 & \bar{j}_1 & \bar{j}_2 & 1 & 1 \\
 j_1+\frac{1}{2} & j_2+\frac{1}{2} & j_1 & j_2 & \frac{1}{2} & \
\frac{1}{2}
\end{array}
\right)_\tmp=-\zfrac{\left(j_1-j_2\right){}^2 \left(2 j_1+2
j_2+3\right) \ \sqrtpw{\left(j_1+j_2+\bar{j}_1-\bar{j}_2+2\right)
\ \left(j_1+j_2-\bar{j}_1+\bar{j}_2+1\right) \
\left(-j_1-j_2+\bar{j}_1+\bar{j}_2\right) \
\left(j_1+j_2+\bar{j}_1+\bar{j}_2+3\right)}}{4
\sqrtpw{\left(j_1+1\right) \left(j_2+1\right)}}
\eed,\bed%
\left(
\begin{array}{@{}c@{\;}c@{}||c@{\;}c@{\;\;}c@{\;}c@{}}
 \bar{j}_1 & \bar{j}_2 & \bar{j}_1 & \bar{j}_2 & 1 & 1 \\
 j_1-\frac{1}{2} & j_2-\frac{1}{2} & j_1 & j_2 & \frac{1}{2} & \
\frac{1}{2}
\end{array}
\right)_\tmp=-\zfrac{\left(j_1-j_2\right){}^2 \left(2 j_1+2
j_2+1\right) \ \sqrtpw{\left(j_1+j_2+\bar{j}_1-\bar{j}_2+1\right)
\ \left(j_1+j_2-\bar{j}_1+\bar{j}_2\right) \
\left(-j_1-j_2+\bar{j}_1+\bar{j}_2+1\right) \
\left(j_1+j_2+\bar{j}_1+\bar{j}_2+2\right)}}{4 \sqrtpw{j_1 j_2}}
\eed,\bed%
\left(
\begin{array}{@{}c@{\;}c@{}||c@{\;}c@{\;\;}c@{\;}c@{}}
 \bar{j}_1 & \bar{j}_2 & \bar{j}_1 & \bar{j}_2 & 1 & 1 \\
 j_1+\frac{1}{2} & j_2-\frac{1}{2} & j_1 & j_2 & \frac{1}{2} & \
\frac{1}{2}
\end{array}
\right)_\tmp=\zfrac{\left(j_1+j_2+1\right){}^2 \left(-2 j_1+2
j_2-1\right) \ \sqrtpw{\left(j_1-j_2+\bar{j}_1-\bar{j}_2+1\right)
\ \left(-j_1+j_2+\bar{j}_1-\bar{j}_2\right) \
\left(j_1-j_2+\bar{j}_1+\bar{j}_2+2\right) \
\left(-j_1+j_2+\bar{j}_1+\bar{j}_2+1\right)}}{4
\sqrtpw{\left(j_1+1\right) j_2}}
\eed,\bed%
\left(
\begin{array}{@{}c@{\;}c@{}||c@{\;}c@{\;\;}c@{\;}c@{}}
 \bar{j}_1 & \bar{j}_2 & \bar{j}_1 & \bar{j}_2 & 1 & 1 \\
 j_1-\frac{1}{2} & j_2+\frac{1}{2} & j_1 & j_2 & \frac{1}{2} & \
\frac{1}{2}
\end{array}
\right)_\tmp=\zfrac{\left(2 j_1-2 j_2-1\right)
\left(j_1+j_2+1\right){}^2 \
\sqrtpw{\left(j_1-j_2+\bar{j}_1-\bar{j}_2\right) \
\left(-j_1+j_2+\bar{j}_1-\bar{j}_2+1\right) \
\left(j_1-j_2+\bar{j}_1+\bar{j}_2+1\right) \
\left(-j_1+j_2+\bar{j}_1+\bar{j}_2+2\right)}}{4 \sqrtpw{j_1
\left(j_2+1\ \right)}}
\eed,\bed%
\left(
\begin{array}{@{}c@{\;}c@{}||c@{\;}c@{\;\;}c@{\;}c@{}}
 \bar{j}_1 & \bar{j}_2 & \bar{j}_1 & \bar{j}_2 & 1 & 1 \\
 j_1 & j_2 & j_1 & j_2 & 0 & 0
\end{array}
\right)_\tmp=-\zfrac{\bar{j}_1^4+4 \bar{j}_1^3+\left(-2
\bar{j}_2^2-2 \ \bar{j}_2+5\right) \bar{j}_1^2+\left(-4
\bar{j}_2^2-4 \bar{j}_2+2\right) \bar{j}_1+\bar{j}_2^4+2
\bar{j}_2^3-5 \left(j_1^2+j_1-j_2 \
\left(j_2+1\right)\right){}^2-\bar{j}_2^2-2
\bar{j}_2}{2 \sqrtpw{5}} \eed%
.

The rest of the reduced coefficients are not listed explicitly
since they can be obtained by using the symmetry properties of the
CG coefficients \cite{R18}. In these remaining cases no
multiplicity occurs,
and the symmetry formula obtains the following form: %
\begin{eqnarray}
\left(\begin{array}{@{}c@{\;}c@{}||c@{\;}c@{\;\;}c@{\;}c@{}}
 \bar{j}_1& \bar{j}_2  & \bar{j'}_1 & \bar{j'}_2 & \bar{1} & \bar{1}\\
 j_1 & j_2 & j'_1 & j'_2 & j''_1 & j''_2
\end{array}\right) &=& (-1)^{\bar{j}_1 - \bar{j'}_1 + \bar{j'}_2 -
\bar{j}_2 + j_1 - j'_1 + j_2 - j'_2 + j''_1 + j''_2} \times  \nonumber \\
& & \sqrt{\frac{dim(\bar{j}_1, \bar{j}_2) (2j'_1 +1)(2j'_2 +
1)}{dim(\bar{j'}_1, \bar{j'}_2) (2j_1 +1)(2j_2 + 1)}}
\left(\begin{array}{@{}c@{\;}c@{}||c@{\;}c@{\;\;}c@{\;}c@{}}
 \bar{j'}_1& \bar{j'}_2  & \bar{j}_1 & \bar{j}_2 & \bar{1} & \bar{1} \\
 j'_1 & j'_2 & j_1 & j_2 & j''_1 & j''_2
\end{array}\right),
\end{eqnarray} %
where $dim(\bar{j}_1, \bar{j}_2) = (2 \bar j_1 - 2 \bar j_2 + 1)
(2 \bar j_1 + 2 \bar j_2 + 3) (2 \bar j_1 + 2) (2 \bar j_2 +
1)/6$.

\section{Conclusion}

In this paper, we present the full accurate list of the CG
coefficients involving the $14$-dimensional irreducible
representation of the $SO(5)$ group, thus correcting numerous
errors in these CG coefficients published previously \cite{R17}.
The CG coefficients listed here are computer-checked in two ways.
Firstly, a Mathematica algorithm is devised that produces
numerical values of the $SO(5)$ CG coefficients. These values were
compared at many points as to coincide with the values evaluated
from the analytic expressions. Secondly, analytic expressions for
the CG coefficients are checked to satisfy the required
orthogonality relations.

\section{Acknowledgments}

This work was supported in part by MNTR, Project-141036. One of us, I.S.,
would like to acknowledge hospitality and useful discussions at the
Institute for Nuclear Research and Nuclear Energy in Sofia (Bulgaria)
during his visit as early stage researcher supported by the FP6 Marie
Curie Research Training Network "Forces-Universe" MRTN-CT-2004-005104.

\end{document}